\documentclass[journal,draftcls,onecolumn,12pt,twoside]{IEEEtranTCOM}
\usepackage{graphicx}
\usepackage{epsfig}
\usepackage{epstopdf}
\usepackage{amstext}
\usepackage{amssymb}
\usepackage{amsmath}
\usepackage{amsthm}
\usepackage[font=footnotesize]{caption}
\usepackage{subcaption}
\usepackage{cite}
\usepackage{float}
\usepackage{array}
\usepackage{url}
\usepackage{flexisym}
\usepackage{soul}
\usepackage[ruled,vlined,linesnumbered]{algorithm2e}

\IEEEoverridecommandlockouts
\DeclareMathOperator*{\argmax}{arg\,max}
\DeclareMathOperator*{\argmin}{arg\,min}

\newcommand\Mark[1]{\textsuperscript#1}

\begin{document}

\title{Full Duplex Operation for Small Cells}
\date{}
\author{\fontsize{11}{13}\selectfont Sanjay Goyal\Mark{1}, Pei Liu\Mark{1}, Shivendra Panwar\Mark{1}, Rui Yang\Mark{2}, Robert A. DiFazio\Mark{2}, Erdem Bala\Mark{2} 
\\\vspace{-1mm}
\Mark{1}New York University Polytechnic School of Engineering, Brooklyn, NY, USA 
\\ \vspace{-2mm}
\Mark{2} InterDigital Communications, Inc., Melville, NY, USA
}

\maketitle
\vspace{-12mm}

\begin{abstract}
Full duplex (FD) communications has the potential to double the capacity of a half duplex (HD) system at the link level. However, FD operation increases the aggregate interference on each communication link, which limits the capacity improvement. In this paper, we investigate how much of the potential doubling can be practically achieved in the resource-managed, small multi-cellular system, similar to the TDD variant of LTE, both in indoor and outdoor environments, assuming FD base stations (BSs) and HD user equipment (UEs). We focus on low-powered small cellular systems, because they are more suitable for FD operation given practical self-interference cancellation limits. A joint UE selection and power allocation method for a multi-cell scenario is presented, where a hybrid scheduling policy assigns FD timeslots when it provides a throughput advantage by pairing UEs with appropriate power levels to mitigate the mutual interference, but otherwise defaults to HD operation. Due to the complexity of finding the globally optimum solution of the proposed algorithm, a sub-optimal method based on a heuristic greedy algorithm for UE selection, and a novel solution using geometric programming for power allocation, is proposed. With practical self-interference cancellation, antennas and circuits, it is shown that the proposed hybrid FD system achieves as much as 94$\%$ throughput improvement in the downlink, and 93$\%$ in the uplink, compared to a HD system in an indoor multi-cell scenario and 53$\%$ in downlink and 60$\%$ in uplink in an outdoor multi-cell scenario. Further, we also compare the energy efficiency of FD operation.

\end{abstract}

\section{Introduction}\label{sec1}
\IEEEPARstart{F}{ull} duplex (FD) operation in a single RF channel can potentially double the spectral efficiency of a wireless network. Approaching this level of improvement poses a number of theoretical and practical challenges but is motivated by the rapid growth in wireless data traffic along with concerns about a spectrum shortage. Regulatory bodies and companies have highlighted these trends with various projections and proposed ways forward~\cite{FCC,Cisco,Ericsson,difazio2011bandwidth,Horizon}. There have even been goals set to improve capacity by as much as 1000x~\cite{EB, Qualcomm}. Recent advances in FD technology~\cite{Khandani10, Katti10, Knox12, Katti13, Duarte13} provide a step towards meeting the projected need without requiring new spectrum. 

The large differential between transmitted (Tx) and received (Rx) powers at a wireless terminal, together with typical Tx/Rx isolation, has driven the vast majority of systems to use either frequency division duplexing (FDD) or time division duplexing (TDD). FDD separates the Tx and Rx signals with filters and TDD with Tx/Rx switching. Recent developments in transceiver design has challenged this limitation, and established the feasibility of FD operation on a common carrier, also known as simultaneous transmit and receive (STR). A combination of antenna, analog and digital cancellation can remove most of the Tx self-interference from the Rx path to allow demodulation of the received signal. This was demonstrated using multiple antennas positioned for optimum cancellation~\cite{Khandani10,Katti10} and later as single antenna systems~\cite{Knox12,Katti13}, where as much as 110 dB cancellation is reported over an 80 MHz bandwidth. Multiple antennas were also used in~\cite{Duarte13}, where the cancellation ranged from 70 to 100 dB with a median of 85 dB. An antenna feed network, for which a prototype provided 40 to 45 dB Tx/Rx isolation before analog and digital cancellation, was described in~\cite{Knox12}. 

Although extensive advances have been made in designing and implementing wireless transceivers with FD capability, and there are some MAC designs for FD IEEE 802.11 systems, to the best of our knowledge, little has been done to understand the impact of such terminals on a cellular network in terms of system capacity and energy efficiency. In~\cite{Duarte13, Sahai11, SanjayAsilomar13, Singh11}, an 802.11 system, with the CSMA/CA MAC slightly modified for FD operation, is presented where software simulations show throughput gains from 1.2x to 2.0x assuming 85 dB cancellation. 

In this paper, we focus on a multi-cellular system, in which only the base stations (BSs) are assumed to be capable of FD operation, where the additional cost and power is most likely to be acceptable, while the user equipment (UE) is limited to half duplex (HD) operation. In such a system, FD operation provides simultaneous uplink and downlink transmission on the same frequency for a pair of UEs. However, while FD operation may increase the capacity by two times, it also generates new intra-cell and inter-cell interference and this is the main challenge we address in this paper.

\begin{figure}
\centering
\includegraphics[width = 3.3in,trim= 0.99in  3.99in 1.58in 4.11in, clip] {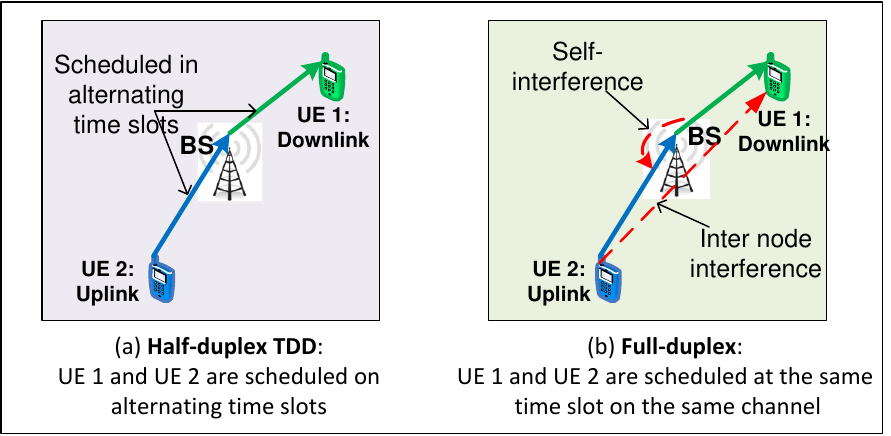}
\vspace{-5mm}
\caption{Half duplex and full duplex single cell scenarios.}
\label{fig:fig1}
\vspace{-5 mm}
\end{figure}

\begin{figure}
\centering
\includegraphics[width = 5in] {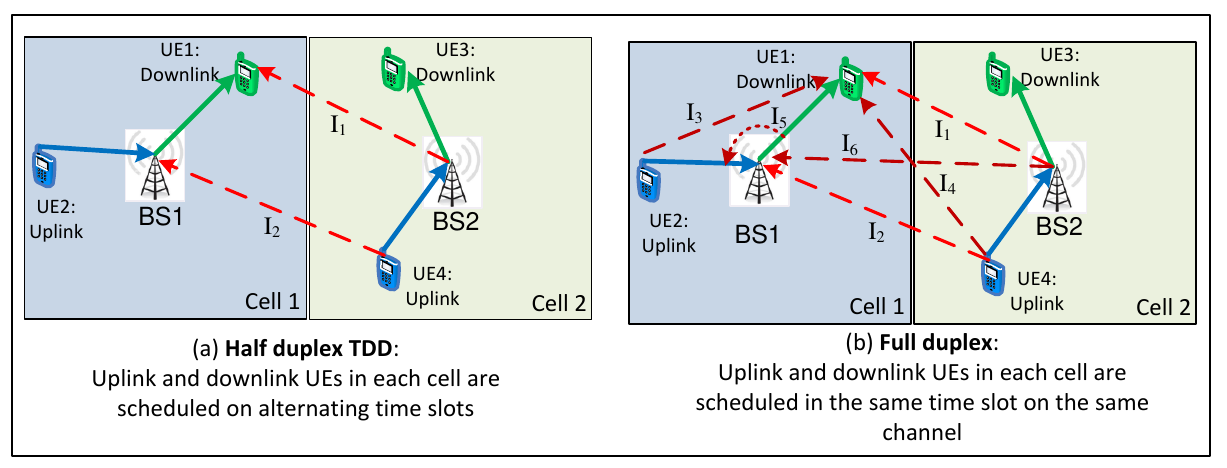}
\vspace{-5mm}
\caption{Half duplex and full duplex multi-cell scenarios.}
\label{fig:fig2}
\vspace{-10 mm}
\end{figure}

The impact of FD operation in a single independent cell and in a multi-cell environment is illustrated in Figure~\ref{fig:fig1} and Figure~\ref{fig:fig2}, respectively. In a single cell, the HD scenario is shown in Figure~\ref{fig:fig1}(a) where UE1 is a downlink UE and UE2 is an uplink UE. The orthogonal channel access in time prevents interference between UEs, but each UE accesses the channel only half the time. Figure~\ref{fig:fig1}(b) shows the FD scenario in a single cell where both UEs are scheduled in the same timeslot, potentially doubling the total cell throughput. Unfortunately, several types of interference which do not exist in the HD scenario need to be considered here: (1) the Tx-to-Rx self-interference at the base station which impacts the ability of the BS to demodulate the uplink signal, and (2) the interference from UE2's uplink signal which impacts the ability of UE1 to demodulate its downlink signal. In a multi-cell scenario, the impact from additional interference during FD operation becomes even more severe because of the inter-cell interference. Consider the two-cell network in Figure~\ref{fig:fig2}, in which UE1 and UE3 are downlink UEs in cell 1 and cell 2, respectively, and UE2 and UE4 are uplink UEs in cell 1 and cell 2, respectively. We assume synchronized cells, which means that in a given time interval all cells schedule transmissions in the same direction. From Figure~\ref{fig:fig2}(a) one can see that in HD operation, UE1 gets interference ($I_1$) from BS2 which is transmitting to UE3 at the same time. Similarly, BS1 gets interference ($I_2$) from the uplink signal of UE4. During FD operation, as shown in Figure~\ref{fig:fig2}(b), the downlink UE, UE1, not only gets interference ($I_1$) from the BS2, but also gets interference ($I_3$ and $I_4$) from the uplink signals of UE2 and UE4. Similarly, the uplink from UE2 to the BS1 not only gets interference ($I_2$) from UE4, but also gets interference ($I_6$) from the downlink signal of  BS2 as well as Tx-to-Rx self-interference ($I_5$). The existence of additional interference sources raises the question of actual gain from FD operation. 

This area has attracted considerable interest. Barghi \emph{et al.}~\cite{Barghi12} compared the capacity of an FD single cell where multiple antennas are used to build an FD radio, to the capacity of a HD single cell where the antennas are used for MIMO transmission. Information theoretic techniques, that is, successive interference cancellation for uplink and dirty paper coding for downlink, are used to calculate the UE capacity. It is shown that under certain conditions, using additional antennas for building an FD radio can provide a performance boost compared to utilizing the antennas to form a high capacity MIMO link. A resource allocation method using matching theory to optimally allocate the subcarriers among Tx-Rx pairs for a single cell FD OFDMA network was proposed by Di \emph{et al.}~\cite{DiINFOCOM}. Shao \emph{et al.}~\cite{Shaocommletter} proposed a cell partitioning method to divide the whole interference region into several partitions and allocate the frequency resources to them for a single cell FD OFDMA system. The methods presented in~\cite{DiINFOCOM} and~\cite{Shaocommletter} cannot be directly applied for resource allocation in a multi-cell scenario. A suboptimal scheduling algorithm to select the transmission direction of each UE in a multi-cell scenario, assuming fixed transmission power for each direction, was proposed by Shen \emph{et al.}\cite{XShen13}. In the FD scenario of \cite{XShen13}, downlink transmission occurs from the center BS, while uplink reception is performed at uniformly distributed Rx antennas. In this system, inter-BS interference and interference from the UEs of neighboring cells is ignored. A similar assumption was made in~\cite{Simeone2014full}, where an analytical expression for the achievable rates assuming Cloud Radio Access Network (C-RAN) operation for both HD and FD are derived. Choi~\cite{HyunICTC} also considered perfect inter-BS interference cancellation while designing a UE pair selection method for the multi-cell FD system. Interference from UEs of neighboring cells were also ignored in \cite{HyunICTC}, which makes the resource allocation easier even for the multi-cell case. However, the assumption of ignoring interference from UEs of neighboring cells may not be appropriate in some scenarios. An cell-edge uplink UE of a neighboring cell may generate severe interference for the downlink transmission. Choi \emph{et al.}~\cite{ChoiSTR12} proposed a method to mitigate the inter-BS interference using null forming in the elevation angle at BS antennas. With this design, they further analyze the performance of the multi-cell system with FD BSs with a simple UE selection procedure by assuming fixed transmission powers in both directions. FD operation in a cellular system has also been investigated in the DUPLO project~\cite{duplo_site}, where a joint uplink-downlink beamforming technique was designed for the single small cell environment~\cite{Nguyenduplo}.

In one of our previous papers~\cite{SanjayCISS13}, we considered a macro multi-cellular network with FD BSs (with complete self-interference cancellation), where an analytical model based on stochastic geometry shows a throughput gain of 11$\%$ and 91$\%$ in the uplink and downlink, respectively. Alves \emph{et al.}~\cite{alves2014average} derived the average spectral efficiency for a stochastic geometry based dense small cell environment with both BS and UEs having FD capability. The throughput gain of a heterogenous network by assuming both BS and UEs with FD capability was derived by Lee \emph{et al.}\cite{Quekhybrid}. They showed the superiority of FD mode for larger access point (AP) densities which contradicts one of the conclusions of this paper. The reason behind this is the lack of inter-BS interference and the approximation of inter-UE interference made in \cite{Quekhybrid}. Only downlink throughput performance is considered in \cite{Quekhybrid}, which does not account for inter-BS interference. In addition, the distance of a UE to a neighboring cell's UE is approximated by the distance from the neighboring cell's AP, resulting in the mitigation of UE to UE interference due to lower UE transmit power. Thus, only self-interference plays a role in the performance difference between FD and HD systems, where by increasing the AP density, BS to UE interference dominates the impact of self-interference. Therefore, HD and FD modes become similar in terms of interference level, which results in higher FD gain due to the higher AP density. Moreover, these stochastic geometry based analyses \cite{SanjayCISS13, alves2014average, Quekhybrid} do not consider multi-UE diversity gain, which comes through scheduling of the appropriate UEs with power adjustments to mitigate interference. This is especially crucial in FD systems, where as we have just noted, the interference scenario is worse than traditional HD systems. In~\cite{SanjayCISS13}, an OFDMA system with a heuristic greedy scheduling algorithm for the UE selection procedure in both FD and HD systems was also simulated, which shows throughput gains of 57$\%$ and 99$\%$ in uplink and downlink, respectively. The design considered the fixed power allocation in both directions and did not consider the effect of residual self-interference at BSs, which is also the case in~\cite{Barghi12, Shaocommletter, XShen13, Simeone2014full, HyunICTC, ChoiSTR12}. Residual self-interference, in general, lowers the uplink coverage and limits the advantage of FD technology in a large cell. For example, consider a cell with a 1 kilometer radius. According to the channel model given in~\cite{3GPP:1}, the path loss at the cell edge is around 130 dB. It means the uplink signal arriving at the BS is 130 dB lower than the downlink signal transmitted, given that equal per channel transmission power in the uplink and downlink directions. The received signal to interference ratio (SIR) is at most -20 dB with the best self-interference cancellation circuit known to date, which is capable of achieving 110 dB of cancellation~\cite{Katti13}. At such an SIR, the spectrum efficiency would be very low. This motivates us to consider small-cell systems as more suitable candidates to deploy an FD BS.

Due to the additional interference sources, the actual gain from FD operation will strongly depend on link geometries, the density of UEs, and propagation effects in mobile channels. Most previous work~\cite{XShen13, Simeone2014full, ChoiSTR12} either ignored or assumed cancellation of strong interference during FD operation. If we do not assume perfect cancellation of strong interference in an FD system, a robust scheduling algorithm is required to intelligently select the UEs with appropriate power levels in all the cells, so that the maximum FD gain can be extracted.  

In prior work~\cite{SanjayICC14}, we set the framework for the single small-cell scenario, where we evaluate link conditions under which FD operation can be supported, and presented a hybrid scheduler that can exploit the FD capability at the BS whenever it is favorable, and otherwise defaults to HD operation. We compared the performance of our hybrid FD scheduler with a HD TDD baseline scheduler by assuming a fixed power allocation per transmission in both the uplink and downlink directions. It was shown by simulation that we can achieve as much as an 81$\%$ increase in capacity (with 85 dB of self-interference cancellation), close to the doubling promised by FD, and we discussed limitations from intra-cell interference effects specific to FD operation.

In this paper, we examine FD common carrier operation applied to a resource managed TDMA-type multi small-cell system for which the TDD variant of LTE is a current example~\cite{DahlmanLTE, 3GPP:2}. In a multi-cell scenario where the interference situation is worse, extracting the throughput gain due to FD operation compared to HD operation is not simple and depends on several factors. It requires an intelligent scheduler which appropriately selects the UEs and their powers during FD operation. We propose a proportional fairness based joint UE selection and power allocation for such a system, to simultaneously select the UEs and transmit power levels to maximize the system gain. This joint UE selection and power allocation is a non-convex, nonlinear, and mixed discrete optimization problem. There exists no method to find a globally optimum solution for such a problem, even for the traditional HD system scenario. We provide a sub-optimal method by separating the UE selection and power allocation procedures, using a heuristic greedy method for UE selection, and using geometric programming for power allocation. For the FD system, the UE selection procedure is a hybrid process, in which FD operation is enabled for a cell where it is advantageous to select two UEs (based on the interference scenario); otherwise it operates in the HD mode. Furthermore, the power allocation procedure adjusts the power of each terminal to an appropriate level so that maximum system gain can be achieved while not violating the maximum power constraint. We compare the performance of FD and HD systems in terms of throughput and energy efficiency in both indoor and outdoor environments by using parameters and simulation assumptions from ongoing activities in the cellular community, for example, 3GPP~\cite{3GPP:1, 3GPP:3}. 

Section~\ref{sec2} presents the FD and HD communication system scenarios in terms of frame structure. Joint UE selection and power allocation algorithms for HD and FD operation are provided in Section~\ref{sec3}. Section~\ref{sec4} contains simulation results for throughput and energy efficiency. Conclusions are discussed in Section~\ref{sec5}.

\section{Full Duplex Frame Structure in a Cellular Environment}\label{sec2}
\begin{figure}
\centering
\includegraphics[width = 4.5in,trim= 1.20in  5in 1.30in 4.7in, clip] {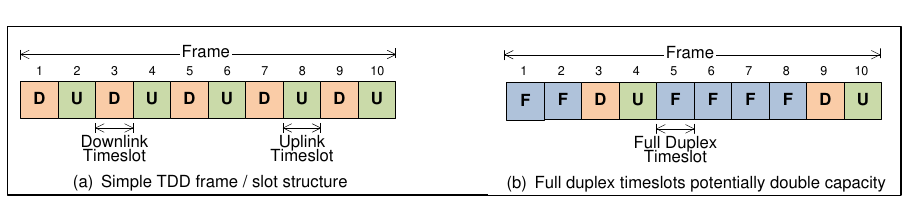}
\vspace{-5mm}
\caption{TDD half duplex baseline and full duplex operation.}
\label{fig:fig3}
\vspace{-10mm}
\end{figure}

We consider a multi-cell deployment scenario in which each cell consists of multiple legacy HD UEs and a BS that can operate in FD or HD mode. Figure~\ref{fig:fig3}(a) shows the frame structure of the HD TDD baseline. It consists of a set of timeslots, all operating on the same frequency channel, that alternate between uplink (U) and downlink (D) operation providing a continuous stream of data in one direction or the other. This is a simplified structure in that a deployed system, TDD LTE for example~\cite{DahlmanLTE, 3GPP:2}, would typically have special timeslots (or subframes) as guard periods for Tx/Rx switching and other overhead functions and may group U and D slots together to minimize switching, which we do not consider in our current analysis. FD timeslots (F) are introduced in Figure~\ref{fig:fig3}(b). It would be desirable to configure every timeslot as FD with the aim of achieving a doubling of capacity, but we anticipate the need to operate some as either solely uplink or solely downlink due to the interference environment explained below. It is the responsibility of a packet scheduler to determine whether a timeslot will be an uplink, downlink, or FD timeslot, and which UE will be given service. 

As was shown in~\cite{SanjayICC14}, the use of FD operation may or may not lead to higher throughput compared to HD operation. The performance of the system depends on multiple factors, such as  the relative locations among UEs and BSs, the propagation channels, the self-interference cancellation capability at the BSs, the required SNR at each receiver, and the Tx power limitations.  Doubling of capacity is only an upper bound, and the actual FD gain needs to be evaluated, which is the subject of the remainder of this paper

\section{UE Selection and Power Allocation}\label{sec3}
As discussed in the previous section that FD throughput gain is available only under certain propagation conditions, distances among nodes in the network, and power levels. This suggests that FD operation should be used opportunistically, that is, with an intelligent scheduler that selects UEs to achieve FD gain when appropriate, and otherwise defaults to HD operation. With this capability, our design of the scheduler attempts to meet the typical criteria of most schedulers: maximize the system throughput while maintaining a level of fairness. In this paper we assume a centralized scheduler that has access to global system information, i.e., channel state information, power, etc. The results generated using this scheduler can be viewed as an upper bound on system performance.

In a multi-cell scenario where each cell consists of multiple UEs, the objective of the scheduler in timeslot $t$ is to maximize the logarithmic sum of the average rates of all the UEs~\cite{Stoylar05}. In this paper, for all systems (HD and FD), we assume that each UE has infinite backlogged data in each direction. In the FD system the scheduler needs to maximize the throughput simultaneously in both uplink and downlink directions. The objective of the scheduler is defined as
\begin{equation}\label{1}
\begin{split}
    & Maximize \sum_{b=1}^B \sum_{k=1}^{N^b} \left[ log(\overline{R^d_{b,k}}(t)) + log(\overline{R^u_{b,k}}(t))\right]  \\[-10pt]
    &\mbox{subject to:} \\[-10pt]
    & \ \ \ \ \ \ \ \  0 \le P^d_{b,k}(t) \le P^{d,max},\\[-10pt]
    & \ \ \ \ \ \ \ \ 0 \le P^u_{b,k}(t) \le P^{u,max},\\[-10pt]
    & \ \ \ \ \ \ \ \ R^d_{b,k}(t) . R^u_{b,k}(t) = 0, \ 1 \le k \le N^b, b = \{1,2,...,B\},
\end{split}
\end{equation} 
where $B$ is the number of cells and  $N^b$ is the number of UEs in cell $b$; $\overline{R^d_{b,k}}(t)$, $\overline{R^u_{b,k}}(t)$ are the average achieved downlink and uplink rates of the UE $k$ in cell $b$, denoted as $UE_{b,k}$, until timeslot $t$, respectively. The first two constraints in (\ref{1}) are for the transmit powers of the UEs and BSs in each cell, in which $P^d_{b,k}(t)$ and $P^u_{b,k}(t)$  are the downlink and uplink transmission powers used in timeslot $t$, corresponding to $UE_{b,k}$, respectively; $P^{d,max}$ and $P^{u,max}$ are the maximum powers that can be used in a downlink and uplink transmission direction, respectively. The third constraint in (\ref{1}) captures the HD nature of the UEs, where $R^d_{b,k}(t) $ and $R^u_{b,k}(t)$ denote the instantaneous rates of $UE_{b,k}$, that can be achieved in timeslot $t$, in the downlink and uplink, respectively. These instantaneous rates are defined later in this section. The average achieved data rate, for example, for downlink, $\overline{R^d_{b,k}}(t)$ is updated iteratively based on the scheduling decision in timeslot $t$, that is, 
\begin{equation}\label{2}
	\overline{R^d_{b,k}}(t) = \left\{ \,
		\begin{IEEEeqnarraybox}[] [c] {l?s}
			\IEEEstrut
			 \beta\overline{R^d_{b,k}}(t-1) + (1-\beta)R^d_{b,k}(t), & if $UE_{b,k}$ is scheduled at timeslot $t$, \\
			\beta\overline{R^d_{b,k}}(t-1),  &  otherwise.
			\IEEEstrut
		\end{IEEEeqnarraybox}
	\right.
 \end{equation}
where $0<\beta<1$ is a constant weighting factor, which is used to calculate the length of the sliding time window, $1/(1-\beta)$, over which the average rate is computed for each frame, and its value is generally chosen close to one, e.g. 0.99~\cite{Stoylar05, jalali2000data, girici2010proportional}. The average achieved uplink rate of $UE_{b,k}$, $\overline{R^u_{b,k}}(t)$ can be similarly defined.

The goal of the scheduler is to select UEs in each cell with appropriate power levels, so that the overall utility defined in (\ref{1}) can be maximized. Assume that $\boldsymbol{\Psi}(t)$ denotes the set of chosen UEs in both downlink and uplink directions in timeslot $t$ as $\boldsymbol{\Psi}(t) = \{\{\psi_1^d(t),\psi_1^u(t)\}, \{\psi_2^d(t),\psi_2^u(t)\},\cdots,\allowbreak \{\psi_B^d(t),\psi_B^u(t)\} \}$. In the $i_{th}$ UE index pair $\{\psi_i^d(t),\psi_i^u(t)\}  (\psi_i^d(t) \neq \psi_i^u(t))$, $\psi_i^d(t)$  is an index of the chosen downlink UE and $\psi_i^u(t)$ is an index of the chosen uplink UE in the $i_{th}$ cell. $\psi_i^d(t) = 0$ ($\psi_i^u(t) = 0$) indicates no UE for the downlink (uplink) in cell $i$. This could be the result of no downlink (uplink) demand in cell $i$, in the current time slot $t$; or, as discussed in the next section, it could also because scheduling any downlink (uplink) transmission in cell $i$, in timeslot $t$ will generate strong interference to the other UEs, and the total network utility will become lower. So, in each timeslot, each cell will select at most one UE in the downlink and at most one UE in the uplink direction. In other words $\psi_i^d(t),\psi_i^d(t) \in \{1,2,\cdots,N^i\}\cup\{0\},\ i = \{1,2,,\cdots,B\}$.

Assume that $\boldsymbol{P}(t) = \left\{ \{p_1^d(t),p_1^u(t)\},\{p_1^d(t),p_1^u(t)\},\cdots,\{p_B^d(t),p_B^u(t)\} \right\}$ contains the power levels for the selected UE combination, $\boldsymbol{\Psi}(t)$, in timeslot $t$, where   $p_i^d(t)$ is the power level of the downlink direction and $p_i^u(t)$ is the power level for the uplink direction in the $i_{th}$ cell. Using (\ref{2}), the objective function in (\ref{1}) can be expressed as
\begin{equation}\label{3}
\begin{split}
 & \textstyle \sum_{b=1}^B \sum_{k=1}^{N^b} [ log(\overline{R^d_{b,k}}(t)) + log(\overline{R^u_{b,k}}(t)) ] = \sum_{b=1}^B [ \{ log(\beta\overline{R^d_{b,\psi_b^d(t)}}(t-1) + (1-\beta)R^d_{b,\psi_b^d(t)}(t)) - \\
 & \textstyle log(\beta\overline{R^d_{b,\psi_b^d(t)}}(t-1)) \} + \{ log(\beta\overline{R^u_{b,\psi_b^u(t)}}(t-1) + (1-\beta)R^u_{b,\psi_b^u(t)}(t)) - log(\beta\overline{R^u_{b,\psi_b^u(t)}}(t-1)) \} ] + A,
 \end{split}
\end{equation}
where $A$ is independent from the decision made at timeslot $t$, and is given by
\begin{equation}\label{4}
A = \sum_{b=1}^B \sum_{k=1}^{N^b} \left[ log(\beta\overline{R^d_{b,k}}(t-1)) + log(\beta\overline{R^u_{b,k}}(t-1))\right].
\end{equation}

In equation (\ref{3}), let us denote the first term in the summation as $\chi^d_{b,\psi_b^d(t)}(t)$,
\begin{equation}\label{5}
\chi^d_{b,\psi_b^d(t)}(t) = log(\beta\overline{R^d_{b,\psi_b^d(t)}}(t-1) + (1-\beta)R^d_{b,\psi_b^d(t)}(t)) - log(\beta\overline{R^d_{b,\psi_b^d(t)}}(t-1)).
\end{equation}
and the second term as $\chi^u_{b,\psi_b^u(t)}(t)$,
\begin{equation}\label{6}
\chi^u_{b,\psi_b^u(t)}(t) = log(\beta\overline{R^u_{b,\psi_b^u(t)}}(t-1) + (1-\beta)R^u_{b,\psi_b^u(t)}(t)) - log(\beta\overline{R^u_{b,\psi_b^u(t)}}(t-1)).
\end{equation}

In the above equations, note that, if $\psi_b^d(t) = 0 \ (\psi_b^u(t) = 0)$, then $\chi^d_{b,\psi_b^d(t)}(t) = 0  \ (\chi^u_{b,\psi_b^u(t)}(t) = 0)$. In the above equations, the instantaneous rates are given by,
\begin{equation}\label{7}
\textstyle R^d_{b,\psi_b^d(t)}(t) = W_c \ log_2(1+{\mathrm{SINR}}_{b,\psi_b^d(t)}) = W_c \ log_2 \left(1 + \frac{p^d_b(t) G_{b,\psi_b^d(t)}}{N_{\psi_b^d(t)} + \sum_{i=1,i \neq b}^B p_i^d(t) G_{i,\psi_b^d(t)} + \sum_{i=1}^B p_i^u(t) G_{\psi_i^u(t),\psi_b^d(t)}}\right),
\end{equation}
\begin{equation}\label{8}
\textstyle R^u_{b,\psi_b^u(t)}(t) = W_c \ log_2(1+{\mathrm{SINR}}_{b,\psi_b^u(t)}) = W_c \ log_2 \left(1 + \frac{p^u_b(t) G_{\psi_b^u(t),b}}{N_{b} + p_b^d(t) \gamma+\sum_{i=1, i \neq b}^B p_i^d(t) G_{i,b} + \sum_{i=1, i \neq b}^B p_i^u(t) G_{\psi_i^u(t),b}}\right).
\end{equation}

In the above equations, $W_c$ is the bandwidth of the channel and G is used to denote the channel gains between different nodes. For example, $G_{b,\psi_b^u(t)}$ denotes the channel gain between BS $b$ and the selected UE $\psi_b^u(t)$; $N_{\psi_b^d(t)}$ and $N_b$  are the noise power at the selected downlink UE and the BS in cell $b$. In (\ref{7}), in denominator of the last term, the second term counts the inter-cell interference from all the other BSs and the third term counts the interference from the uplink UEs of all cells. In (\ref{8}), in denominator of the last term, the second term counts the self-interference at its own BS, where $\gamma$ is used to denote the self interference cancellation level at the BS; the third term counts the inter-cell interference from the BSs of other cells; and the fourth term includes the inter-cell interference from uplink UEs of other cells.

The overall utility of a cell (e.g. cell $b$) is defined as
\begin{equation}\label{9}
\Phi_{b,(\psi_b^d(t),\psi_b^u(t))}(t) = \chi^d_{b,\psi_b^d(t)}(t) + \chi^u_{b,\psi_b^u(t)}(t);
\end{equation}

Then, the optimization problem in (\ref{1}) can be equivalently expressed as
\begin{equation}\label{10}
\begin{split}
    & \argmax_{(\boldsymbol{\Psi}(t), \boldsymbol{P}(t))}  \sum_{b=1}^B \Phi_{b,(\psi_b^d(t),\psi_b^u(t))}(t)\\
    &\mbox{subject to:} \\[-10pt]
    & \ \ \ \ \ \ \ \ 0 \le p^d_{b}(t) \le P^{d,max},\\[-10pt]
    & \ \ \ \ \ \ \ \ 0 \le p^u_{b}(t) \le P^{u,max},\\[-10pt]
    & \ \ \ \ \ \ \ \ \psi_b^d(t) \neq \psi_b^u(t), \  b = \{1,2,...,B\},
\end{split}
\end{equation}

The above problem is a nonlinear nonconvex combinatorial optimization and computing its globally optimal solution may not be feasible in practice. Although the problem can be optimally solved via exhaustive search, the complexity of this method increases exponentially as the number of cells increase. Moreover, the above problem is a mixed discrete (UE selection) and continuous (power allocation) optimization. In this paper, a joint UE selection and power allocation is proposed, which achieves near-optimal solution through iterative algorithms. 

We solve the joint UE selection and power allocation problem (\ref{10}) in each timeslot in two steps, (1) $\textit  {UE Selection}$: for a given feasible power allocation, this step finds the UE combination with maximum overall utility, and (2) $\textit {Power Allocation}$: for the given UE combination, this step derives the powers to be allocated to the selected UEs such that overall utility can be maximized. In the next two subsections, we discuss both steps in detail. 
\subsection{UE Selection }\label{sec3a}
In this step, for each timeslot $t$, for the given power allocation ($\boldsymbol {P}_{initial}(t)$ ), the objective of the centralized scheduler is to find the UEs in each cell to transmit, which is given as 

\begin{equation}\label{11}
\begin{split}
    & \boldsymbol{\Psi}^*(t) =\argmax_{\boldsymbol{\Psi}(t)}  \sum_{b=1}^B \Phi_{b,(\psi_b^d(t),\psi_b^u(t))}(t)\\
    &\mbox{subject to:} \\[-10pt]
    & \ \ \ \ \ \ \ \ \psi^d_{b}(t) \neq \psi^u_{b}(t), \  b = \{1,2,...,B\}.
\end{split}
\end{equation} 

In the above problem, the constraint captures the HD nature of the UEs, which is similar to the third constraint in the problem formulation (\ref{1}). 

In the traditional HD systems, finding the optimal set of UEs is very different in the downlink and uplink direction. In the literature, the problem above is solved optimally in the downlink direction~\cite{venturino2009coordinated, yu2010joint, kiani2007maximizing}, where the interferers are the fixed BSs (assuming a synchronized HD multi-cell system) in the neighboring cells. It is easy to estimate the channel gains for each UE with the neighboring BSs. Thus, interference from the neighboring cells can be calculated without knowing the actual scheduling decision (UE selection) of the neighboring cells. In this situation, a centralized scheduler can calculate the instantaneous rate and the utility of the each UE in the each cell, and make the UE selection decision for each cell optimally. In the uplink scheduling, for the given power allocation, interference from the neighboring cell cannot be calculated until the actual scheduling decision of the neighboring cell is known, because in this case, a UE in the neighboring cell generates the interference. This is also applied to the FD system, where interference from the neighboring cell could be from a UE or the BS or both. 

To solve this problem, we use a heuristic method similar to~\cite{SanjayCISS13,koutsopoulos2006cross}. We provide a centralized greedy algorithm to achieve a sub-optimal solution. The algorithm runs at the start of each timeslot, which we call Algorithm 1.

\begin{algorithm}
\small {
$\mathbf{Q}_B$ = 0; $\mathbf{R}_{B}$ = 0\;\vspace{-2mm}
$\Theta$ = A random order of the sequence of all the BSs\;\vspace{-2mm}
\For{c = $\Theta$(1) to $\Theta$(B)} {\vspace{-2mm}
$\alpha_c^d = \{1,2,\cdots,N^c\}$, $\alpha_c^u = \{1,2,\cdots,N^c\}$\;\vspace{-2mm}
 $\left\{\psi_c^d(t), \Delta U_c^d(t) \right\} =\left\{ \argmax_{d \in \alpha_c^d} \{Get\_Utility(c,d,0)\}, Get\_Utility(c,\psi_c^d(t),0)\right\}$\;
 $\left\{\psi_c^u(t), \Delta U_c^u(t) \right\} =\left\{ \argmax_{u \in \alpha_c^u} \{Get\_Utility(c,0,u)\}, Get\_Utility(c,0,\psi_c^u(t))\right\}$\;
\If{$\Delta U_c^d(t) > \Delta U_c^u(t)$ and $\Delta U_c^d(t) > 0$} {\vspace{-2mm}
$\mathbf{R} \leftarrow$ set $R(c) = \psi_c^d(t)$ in $\mathbf{R}$\;\vspace{-2mm}
}
\ElseIf{$\Delta U_c^u(t) > \Delta U_c^d(t)$ and $\Delta U_c^u(t) > 0$} {\vspace{-2mm}
$\mathbf{Q} \leftarrow$ set $Q(c) = \psi_c^u(t)$ in $\mathbf{Q}$\;\vspace{-2mm}
}
}\vspace{-2mm}
\For{c = $\Theta$(1) to $\Theta$(B)} {\vspace{-2mm}
\If{$R(c) \neq 0$}{\vspace{-2mm}
$\left\{\psi_c^u(t), \Delta U_c^u(t) \right\} =\left\{ \argmax_{u \in \alpha_c^u\backslash R(c)} \{Get\_Utility(c,R(c),u)\}, Get\_Utility(c,R(c),\psi_c^u(t))\right\}$\;
\If {$\Delta U_c^u(t) > 0$} {\vspace{-2mm}
$\mathbf{Q} \leftarrow$ set $Q(c) = \psi_c^u(t)$ in $\mathbf{Q}$\;\vspace{-2mm}
}
}\vspace{-2mm}
\ElseIf {$Q(c) \neq 0$} {
$\left\{\psi_c^d(t), \Delta U_c^d(t) \right\} =\left\{ \argmax_{d \in \alpha_c^d\backslash Q(c)} \{Get\_Utility(c,d,Q(c))\}, Get\_Utility(c,\psi_c^d(t),Q(c))\right\}$\;
\If {$\Delta U_c^d(t) > 0$} {\vspace{-2mm}
$\mathbf{R} \leftarrow$ set $R(c) = \psi_c^d(t)$ in $\mathbf{R}$\;\vspace{-2mm}
}}}}
\label{alg1}
\caption{UE Selection ($\boldsymbol{P}_{initial}(t)$)}
\end{algorithm}

In each timeslot, the algorithm first initializes the vectors that contain the allocation results. Vectors $\mathbf{Q}$ and $\mathbf{R}$ contain the information of scheduled uplink UEs and downlink UEs, respectively, which are iteratively updated as the scheduling decision is taken for a cell. The entry $Q(i)$ in $\mathbf{Q}$ contains the index of scheduled uplink UE of BS $i$, if any, otherwise it will be zero. Similarly, entry $R(i)$of matrix $\mathbf{R}$ contains the index of the scheduled dowlink UE in cell $i$, if any, otherwsie zero. Note that in any timeslot, $Q(i) \neq R(j)$, if $i = j$ and $Q(i) \neq 0, R(j) \neq 0$ to ensure the HD constraint for UEs. In each timeslot $t$, the centalized scheduler generates a random order of the BSs (Line 2). Following that given order of the BSs, in each cell, the algorithm first finds the UE with the maximum positive utility gain, which can be either in the uplink direction (i.e., $\psi_c^u(t)$ for cell $c$ ) or in the downlink direction (i.e., $\psi_c^d(t)$ for cell $c$ ) (Line 4 - Line 10). To calculate the utility gain in each case, it uses a function $Get\_Utility(.)$ given in Algorithm 2, which is discussed later. It also updates the vector $\mathbf{Q}$ or $\mathbf{R}$ based on the decision made (Line 7 - Line 10). Now, to use the FD capability of the BS, the algorithm again runs for the same order of the BSs (Line 11 - Line 19). For each BS, it finds the UE with the maximum positive utility gain in the opposite direction of what has been selected in the previous loop (if any). Finally, based on the decision, it also updates the vector $\mathbf{Q}$ or $\mathbf{R}$. 
\begin{algorithm}
\small{
\If{$Q(c) = 0$ and $u \neq 0$}{\vspace{-2mm}
$\mathbf{Q^{'}} \leftarrow$ set $Q(c) = u$ in $\mathbf{Q}$\;\vspace{-2mm}
$\mathbf{R^{'}} \leftarrow$ $\mathbf{R}$\;\vspace{-2mm}
$U_{gain} \leftarrow U(u,c,\mathbf{Q^{'}},\mathbf{R^{'}})$\;\vspace{-2mm}
$U_{loss\_uplink} \leftarrow \sum_{i=Q(k):i\neq0, \forall k \in \Theta \backslash c} \left\{ U(i,k,\mathbf{Q^{'}},\mathbf{R^{'}}) - U(i,k,\mathbf{Q},\mathbf{R}) \right\}$\;
$U_{loss\_downlink} \leftarrow \sum_{i=R(k):i\neq0, \forall k \in \Theta} \left\{ U(i,k,\mathbf{Q^{'}},\mathbf{R^{'}}) - U(i,k,\mathbf{Q},\mathbf{R}) \right\}$\;
}
\ElseIf{$R(c) = 0$ and $d \neq 0$}{\vspace{-2mm}
$\mathbf{R^{'}} \leftarrow$ set $R(c) = d$ in $\mathbf{R}$\;\vspace{-2mm}
$\mathbf{Q^{'}} \leftarrow$ $\mathbf{Q}$\;\vspace{-2mm}
$U_{gain} \leftarrow U(d,c,\mathbf{Q^{'}},\mathbf{R^{'}})$\;\vspace{-2mm}
$U_{loss\_uplink} \leftarrow \sum_{i=Q(k):i\neq0, \forall k \in \Theta} \left\{ U(i,k,\mathbf{Q^{'}},\mathbf{R^{'}}) - U(i,k,\mathbf{Q},\mathbf{R}) \right\}$\;
$U_{loss\_downlink} \leftarrow \sum_{i=R(k):i\neq0, \forall k \in \Theta \backslash c} \left\{ U(i,k,\mathbf{Q^{'}},\mathbf{R^{'}}) - U(i,k,\mathbf{Q},\mathbf{R}) \right\}$\;
}
$\Delta U = U_{gain} - |U_{loss\_uplink}| - |U_{loss\_downlink}|$\;\vspace{-2mm}
return $\Delta U$;
}
\label{alg2}
\caption{$Get\_Utility(c,d,u)$}
\end{algorithm}
\setlength{\textfloatsep}{0pt}%

Next, we describe how the function $Get\_Utility(.)$ works. As shown in Algorithm 2, it calculates the utility gain $\Delta U$ for the given cell and the UE based on the transmission direction, i.e., either in the uplink (Line 1- Line 6) or in the downlink  (Line 7- Line 12). The utility gain $\Delta U$ is the difference between the gain in the marginal utility of the chosen UE ($U_{gain}$) and loss in the marginal utility of other uplink and downlink UEs ( $|U_{loss\_uplink}|$ and $|U_{loss\_downlink}|$) due to new interference generated from the chosen UE. Since, in this algorithm, the channel is allocated sequentially cell by cell, thus, $U_{gain}$ is the gain in utility due to scheduling of UE $i$ (say for BS $c$ and slot $t$), which is given by $U(i,c,\mathbf{Q^'},\mathbf{R^'})$, and it is calculated using (\ref{5}) for downlink or (\ref{6}) for uplink, by considering the channel allocation according to $\mathbf{Q^'}$ and $\mathbf{R^'}$. It means that during the calculation of the instantaneous rates in (\ref{5}) or (\ref{6}), it only considers the interference from the cells in which channel has been already assigned, which is given in $\mathbf{Q^'}$ and $\mathbf{R^'}$. Similarly, the utility loss for UEs, to which channel has been already assigned, is calculated as the difference in utility with the new interference occurring due to scheduling of new UEs and without this interference. Equations (\ref{5}) and (\ref{6}) are used to calculate both marginal utility terms, i.e., with and without new interference for $U_{loss\_uplink}$ and $U_{loss\_downlink}$, respectively.

\begin{figure}
\centering
\includegraphics[width = 2.5in,trim= 2.02in  4.49in 2.14in 4.24in, clip] {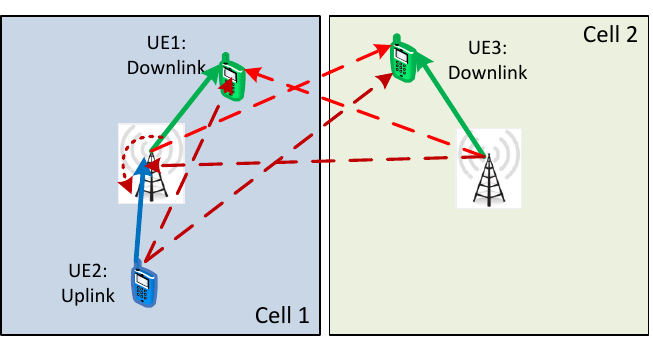}
\caption{An example of partial full duplex operation, where cell 2 is in half duplex mode and cell 1 is in full duplex mode.}
\label{fig:fig4}
\end{figure}
Algorithm 1 gives the UE combination $\boldsymbol{\Psi}^*(t)$, as $\boldsymbol{\Psi}^*(t) = \{\{R(1),Q(1)\}, \{R(2),Q(2)\},\cdots,\allowbreak \{R(B),Q(B)\} \}$. It consists of a downlink UE, or an uplink UE, or both, or no UE from each cell. It is a hybrid FD scheduling algorithm, where, in each timeslot, a cell can be in FD operation, or in HD operation, or no operation at all. An example is given in Figure~\ref{fig:fig4} for two cells, where cell 1 is in FD operation and cell 2 is in HD operation. To evaluate the performance of the FD system, we use an HD system as the benchmark, in which we assume that the transmission direction (uplink or downlink) of all cells are synchronized and follows the frame structure shown in Figure~\ref{fig:fig3}(a). For the HD system, we also use the same procedure for UE selection. In each timeslot, for example, for uplink, we apply the same algorithm as discussed above and find the UE combinations consisting of an uplink UE or no UE from each cell. In the next subsection, we discuss the power allocation procedure for the selected UEs.

\subsection{Power Allocation }\label{sec3b}
In this step, for the selected UE combination in the step 1, a power allocation process is applied to find the appropriate power levels for all UEs, so that the overall utility can be maximized, or for the given $\boldsymbol{\Psi}^*(t)$ from the previous step (note that in this subsection, we use $\boldsymbol{\Psi}(t)$ to denote $\boldsymbol{\Psi}^*(t)$, which is the UE selection found in the previous subsection),
\begin{equation}\label{12}
\begin{split}
    & \boldsymbol{P}^*(t) = \argmax_{\boldsymbol{P}(t)}  \sum_{b=1}^B \Phi_{b,(\psi_b^d(t),\psi_b^u(t))}(t)\\
    &\mbox{subject to:} \\[-10pt]
    & \ \ \ \ \ \ \ \ 0 \le p^d_{b}(t) \le P^{d,max},\\[-10pt]
    & \ \ \ \ \ \ \ \ 0 \le p^u_{b}(t) \le P^{u,max}, \  b = \{1,2,...,B\}.
\end{split}
\end{equation} 

The above optimization is also a nonlinear nonconvex problem, which does not have any method for a low complexity solution. To get a near-optimal solution, we use geometric programming (GP)~\cite{boyd2007tutorial, chiang2007power}. GP cannot be applied directly to the objective function given in (\ref{12}) so we first convert our objective function into a weighted sum rate maximization using approximations as described below. 

In (\ref{12}), the aggregate utility $\Phi_{b,(\psi_b^d(t),\psi_b^u(t))}(t)$ is the sum of the downlink and uplink UE's utility. Let us consider the downlink utility term to show the simplification procedure; the same procedure can be directly applied to the uplink utility term. For example, consider the downlink utility as given in (\ref{5}). It can also be written as,
\begin{equation}\label{13}
\chi^d_{b,\psi_b^d(t)}(t) = log\left( 1+ \frac{(1-\beta)R^d_{b,\psi_b^d(t)}(t)}{\beta\overline{R^d_{b,\psi_b^d(t)}}(t-1)}\right).
\end{equation}

In the above equation, $\beta \in (0,1)$ with a value close to one (e.g. $\beta$ =0.999, or 0.99)~\cite{jalali2000data, girici2010proportional}. Moreover, if we assume that the value of the instantaneous rate, $R^d_{b,\psi_b^d(t)}$, will be close to the average rate, $\overline{R^d_{b,\psi_b^d(t)}}$, then the term $\frac{(1-\beta)R^d_{b,\psi_b^d(t)}(t)}{\beta\overline{R^d_{b,\psi_b^d(t)}}(t-1)}$ will be close to zero. So, by using $ln(1+x) \approx x$ for $x$ close to zero, (13) can be converted to,
\begin{equation}\label{14}
\chi^d_{b,\psi_b^d(t)}(t) \simeq  w_{b,\psi_b^d(t)} R^d_{b,\psi_b^d(t)}(t),
\end{equation}
where, the weight of the UE $\psi_b^d(t)$ is given by,
\begin{equation}\label{15}
w_{b,\psi_b^d(t)} = \frac{(1-\beta)}{\beta \overline{R^d_{b,\psi_b^d(t)}}(t-1)}.\frac{1}{ln(10)}
\end{equation}
Thus, the problem (\ref{12}) can be converted to,
\begin{equation}\label{16}
\begin{split}
    & \boldsymbol{P}^*(t) = \argmax_{\boldsymbol{P}(t)}  \sum_{b=1}^B w_{b,\psi_b^d(t)} R^d_{b,\psi_b^d(t)}(t) +  \sum_{b=1}^B w_{b,\psi_b^u(t)} R^u_{b,\psi_b^u(t)}(t)\\
    &\mbox{subject to:} \\[-10pt]
    & \ \ \ \ \ \ \ \ 0 \le p^d_{b}(t) \le P^{d,max},\\[-10pt]
    & \ \ \ \ \ \ \ \ 0 \le p^u_{b}(t) \le P^{u,max}, \  b = \{1,2,...,B\}.
\end{split}
\end{equation} 
which can be further written as,
\begin{equation}\label{17}
\begin{split}
    & \argmin_{\boldsymbol{P}(t)} {\prod_{b=1}^B \left( \left( \frac{1}{1+{\mathrm{SINR}}_{b,\psi_b^d(t)}}\right)^{w_{b,\psi_b^d(t)}}. \left( \frac{1}{1+{\mathrm{SINR}}_{b,\psi_b^u(t)}}\right)^{w_{b,\psi_b^u(t)}} \right)}\\
    &\mbox{subject to:} \\[-10pt]
    & \ \ \ \ \ \ \ \ 0 \le \frac{p^d_{b}(t)}{P^{d,max}} \le 1,\\[-3pt]
    & \ \ \ \ \ \ \ \ 0 \le \frac{p^u_{b}(t)}{P^{u,max}} \le 1, \  b = \{1,2,...,B\}.
\end{split}
\end{equation} 

In general, to apply GP, the optimization problem should be in GP standard form~\cite{boyd2007tutorial, chiang2007power}. In the GP standard form, the objective function is a minimization of a $\textit {posynomial}$\footnote{ A monomial is a function $f:\mathbf{R}_{++}^n \rightarrow \mathbf{R}: g(p) = d p_1^{a^{(1)}}p_2^{a^{(2)}}\cdots p_n^{a^{(n)}}$, where $d \geq 0$ and $a^{(k)} \in \mathbf{R}, k = 1,2,\cdots,n.$ A posynomial is a sum of monomials, $f(p) = \sum_{j=1}^J d_j p_1^{a_j^{(1)}} p_2^{a_j^{(2)}} \cdots p_n^{a_j^{(n)}}$. } function; the inequalities and equalities in the constraint set are a posynomial upper bound inequality and $\textit {monomial}$ equality, respectively. 

In our case, in (\ref{17}), constraints are monomials (hence posynomials), but the objective function is a ratio of posynomials, as shown in  (\ref{18}). Hence, (\ref{17}) is not a GP in standard form, because posynomials are closed under multiplication and addition, but not in division.
\begin{equation}\label{18}
\begin{split}
& \textstyle {\prod_{b=1}^B \left( \left( \frac{1}{1+{\mathrm{SINR}}_{b,\psi_b^d(t)}}\right)^{w_{b,\psi_b^d(t)}}. \left( \frac{1}{1+{\mathrm{SINR}}_{b,\psi_b^u(t)}}\right)^{w_{b,\psi_b^u(t)}} \right) =} \\
&\scriptstyle { \prod_{b=1}^B \left( \left( \frac{N_{\psi^d_b(t)} + \sum_{i=1, i \neq b}^B p_i^d(t) G_{i,\psi^d_b(t)} + \sum_{i=1}^B p_i^u(t) G_{\psi^u_i(t),\psi^d_b(t)}}{N_{\psi^d_b(t)} + \sum_{i=1}^B p_i^d(t) G_{i,\psi^d_b(t)} + \sum_{i=1}^B p_i^u(t) G_{\psi^u_i(t),\psi^d_b(t)}}\right)^{w_{b,\psi_b^d(t)}}. \left( \frac{N_b + p_b^d(t) \gamma +  \sum_{i=1, i \neq b}^B p_i^d(t) G_{i,b} + \sum_{i=1, i \neq b}^B p_i^u(t) G_{\psi^u_i(t),b}}{N_b + p_b^d(t) \gamma +  \sum_{i=1, i \neq b}^B p_i^d(t) G_{i,b} + \sum_{i=1}^B p_i^u(t) G_{\psi^u_i(t),b}}\right)^{w_{b,\psi_b^u(t)}} \right)}
\end{split}
\end{equation}
 
According to~\cite{chiang2007power}, (\ref{17}) is a signomial programming (SP) problem. In~\cite{chiang2007power}, an iterative procedure is given, in which (\ref{17}) is solved by constructing a series of GPs, each of which can easily be solved. In each iteration of the series, the GP is constructed by approximating the denominator posynomial (\ref{18}) by a monomial, then using the arithmetic-geometric mean inequality and the value of $\boldsymbol{P}$ from the previous iteration. The series is initialized by any feasible $\boldsymbol{P}$, and the iteration is terminated at the $s_{th}$ loop if $||\boldsymbol{P}_s - \boldsymbol{P}_{s-1}|| < \epsilon $, where $\epsilon$ is the error tolerance. This procedure is provably convergent, and empirically almost always computes the optimal power allocation~\cite{chiang2007power}. 

In the overall joint UE selection and power allocation procedure as shown in the Algorithm~3, for each timeslot, we start with maximum capability of UEs (i.e., maximum powers) for each direction to perform the UE selection procedure as given in the last subsection, which provides the UE combination to be scheduled. Then, in second step, the power allocation process, as discussed above, is applied for this given UE combination to find the optimum powers for selected UEs. In the case, when no feasible power allocation for the selected UE combination is found from the power allocation process, a UE with the lowest utility gain is removed from the combination, followed by again applying the power allocation procedure. This process is continued until the feasibility issue is resolved.
\begin{algorithm}
\small{
$\boldsymbol{P}_{initial}(t) = \left\{ \{P^{d,max}, P^{u,max}\}, \{P^{d,max}, P^{u,max}\}, \cdots,\{P^{d,max}, P^{u,max}\} \right\}$\;\vspace{-2mm}
$\boldsymbol{\Psi}^*(t)$ = UE Selection ($\boldsymbol{P}_{initial}(t)$)\;\vspace{-2mm}
loop: \\\vspace{-2mm}
\If {Solution(Geometric Programming($\boldsymbol{\Psi}^*(t)$)) is feasible} {\vspace{-2mm}
$\boldsymbol{P}^*(t)$ = Geometric Programming($\boldsymbol{\Psi}^*(t)$))\;\vspace{-2mm}
}\vspace{-2mm}
\Else{\vspace{-2mm}
$\theta(t)$ = UE with the lowest utility gain\;\vspace{-2mm}
$\boldsymbol{\Psi}^*(t)$) = $\boldsymbol{\Psi}^*(t)) \backslash \theta(t)$\;\vspace{-2mm}
goto loop;\vspace{-2mm}
}
}\label{alg3}
\caption{Overall Joint Selection and Power Control}
\end{algorithm}
\setlength{\textfloatsep}{0pt}%

To generate the results for the HD base system, we use the same procedure in each timeslot in the corresponding direction. For example, in this case, (\ref{16}), (\ref{17}), and (\ref{18}) will just contain the single term for the corresponding direction in place of two terms.

\section{Performance Evaluation}\label{sec4}
\begin{figure}
\centering
\includegraphics[width = 6in] {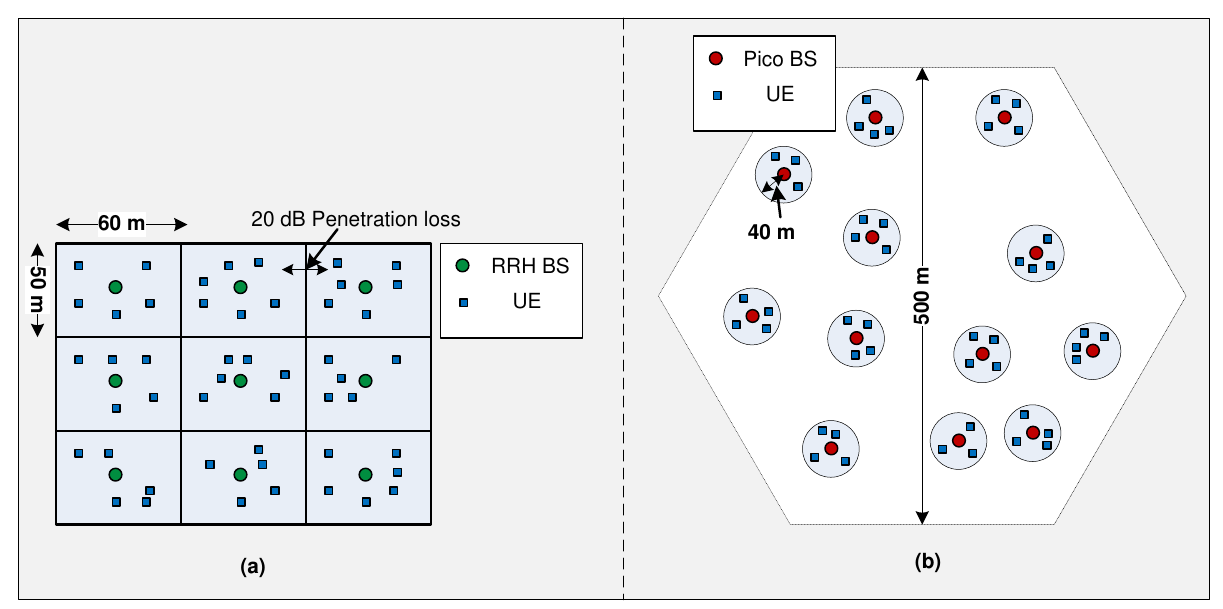}
\caption{(a) An indoor environment with nine RRH Cells, (b) An outdoor environment with twelve Pico cells.}
\label{fig:fig5}
\end{figure}

In this section, we present a simulation analysis comparing the throughput and energy efficiency of the FD and the HD systems using the joint UE selection and power allocation algorithm described in Section~\ref{sec3}. Two deployment scenarios are studied: a dense indoor multi-cell system with nine indoor Remote Radio Head (RRH)/Hotzone cells, as shown in Figure~\ref{fig:fig5}(a), and a sparse outdoor multi-cell system with twelve randomly dropped Pico cells, as shown in Figure~\ref{fig:fig5}(b). As we described in Section~\ref{sec1}, since FD operation increases the interference in a network significantly, we select these two particular small cell scenarios to analyze the performance of FD operation because the penetration loss between cells in the indoor environment, and sparsity in the outdoor environment, provides some static relief in inter-cell interference. The channel bandwidth is 10 MHz for both the HD and the FD systems in both scenarios. In our simulations, since we use the Shannon equation to measure the data rate, we apply a minimum spectral efficiency of 0.26 bits/sec/Hz and a maximum spectral efficiency of 6 bits/sec/Hz to match practical systems. BSs and UEs are assumed to be equipped with single antennas. All other simulation parameters for each scenario are defined below in its corresponding sub-section.

\subsection{Simulation results for dense indoor multi-cell environment}\label{sec4a}

In this section we present the results for the dense indoor multi-cell environment as shown in Figure~\ref{fig:fig5}(a). The simulation parameters, based on 3GPP simulation recommendations for an RRH cell environment~\cite{3GPP:3}, are described in Table~\ref{tab1}. Path loss for both $LOS$ and $NLOS$ within a cell are given in Table~\ref{tab1}, where the probability of $LOS$ ($P_{LOS}$) is,
\begin{equation}\label{19}
\small{
	P_{LOS} = 
	\left \{ \,
		\begin{IEEEeqnarraybox}[] [c] {l?s}
			\IEEEstrut
			1 &  $R \leq 0.018$, \\[-10pt]
			\exp{(-(R-0.018)/0.027)} & $ 0.018 < R < 0.037$, \\[-10pt]
			0.5 & $R \geq 0.037$,
			\IEEEstrut
		\end{IEEEeqnarraybox}  
	\right. 
	}
\end{equation}

In (\ref{19}), $R$ is the distance in kilometers. The channel model used between BSs and UEs is also used between mobile UEs and between BSs for the FD interference calculations with the justification that BSs do not have a significant height advantage in the small cell indoor scenarios considered, and that it is a conservative assumption for the UE-to-UE interference channel. Eight randomly distributed UEs are deployed in each cell. With these settings, we run our simulation for different UE drops in all cells, each with a thousand timeslots, with the standard wrap around topology and generate results for both the HD and FD systems. 

\begin {table}
\caption {Simulation parameters for indoor multi-cell scenario} \label{tab1} 
\begin{center}
    \begin{tabular}{| p{2.7 in} | p{3.4 in} |}
    	\hline
		\textbf {Parameter} & \textbf{Value} \\ \hline
		Maximum BS Power & 24 dBm \\ \hline
		Maximum UE Power & 23 dBm \\ \hline
		Thermal Noise Density & -174 dBm/Hz \\ \hline
		Noise Figure & BS: 8 dB, UE: 9 dB \\ \hline
		Shadowing standard deviation (with no correlation) &  LOS:~ $3~dB$ NLOS: $4~dB$ \\  \hline
		Path Loss within a cell (dB) (R in kilometers)&  LOS: $89.5 + 16.9~log_{10}(R)$, 
		NLOS: $147.4 + 43.3~log_{10}(R)$ \\ \hline
		Path Loss between two cells (R in kilometers)& Max(($131.1 + 42.8~log_{10}(R)),(147.4 + 43.3~log_{10}(R)))$ \\ \hline
		Penetration loss& Due to boundary wall of an RRH cell: 20 dB, Within a cell: 0 dB \\ \hline
    \end{tabular}
\end{center}
\end{table}

\begin{figure}
\centering
\includegraphics[width = 5in] {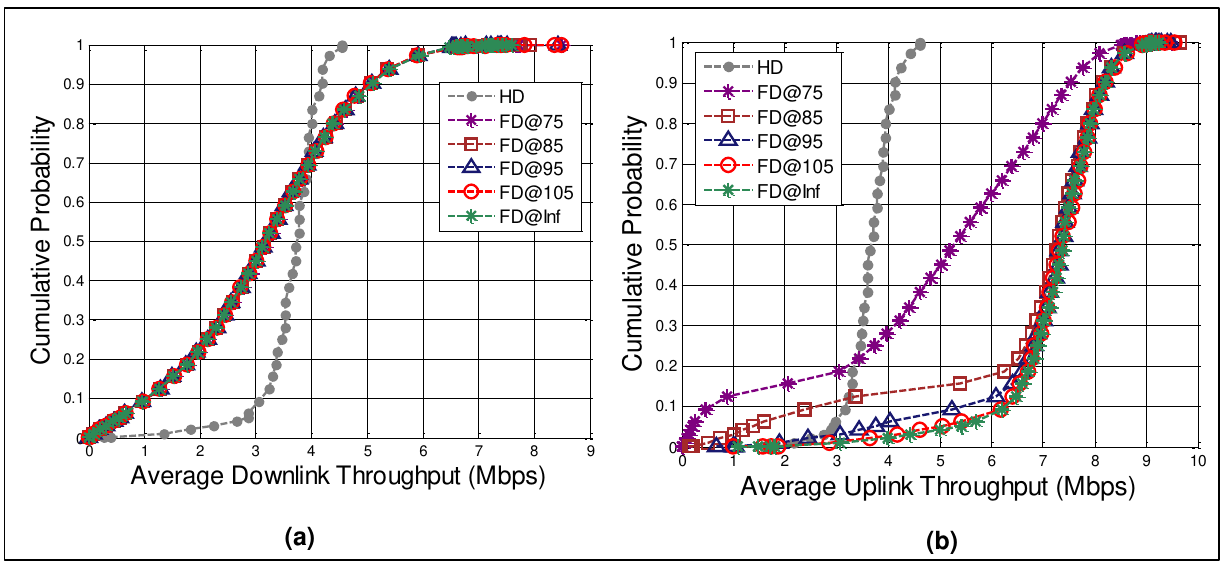}
\caption{Distribution of average data rates for the half-duplex system and full-duplex system with round-robin scheduler in indoor multi-cell scenario.}
\label{fig:fig6}
\end{figure}

We first generate the results for a round-robin scheduler with fixed transmission powers, that is, maximum allowed power in both directions. In the HD system, in each direction, each cell selects UEs in the round-robin manner. In the FD system, in each timeslot, each cell chooses the same UE as selected in the HD system with a randomly selected UE for the other direction to make an FD pair. Figures~\ref{fig:fig6}(a) and~\ref{fig:fig6}(b) show the distribution of average downlink and uplink throughputs, for the different BS self-interference cancellation capability. \emph{FD@x} means the FD system with self-interference cancellation of $x$ dB. \emph{FD@Inf} means that there is no self-interference. In the downlink direction, in most of the cases (~70$\%$), there is no FD gain, which is due to the lack of any intelligent selection procedure during FD operation. In the uplink, due to the cancellation of self-interference, the FD system has a gain compared to the HD system, which increases with the self-interference cancellation. From a complete system point of view, which includes both uplink and downlink, this round-robin scheduling does not provide FD capacity gain in most of the cases. This demonstrates the need for an intelligent scheduling algorithm to provide gain during FD operation, which can benefit both uplink and downlink. 

\begin{figure}
\centering
\includegraphics[width = 5.5in] {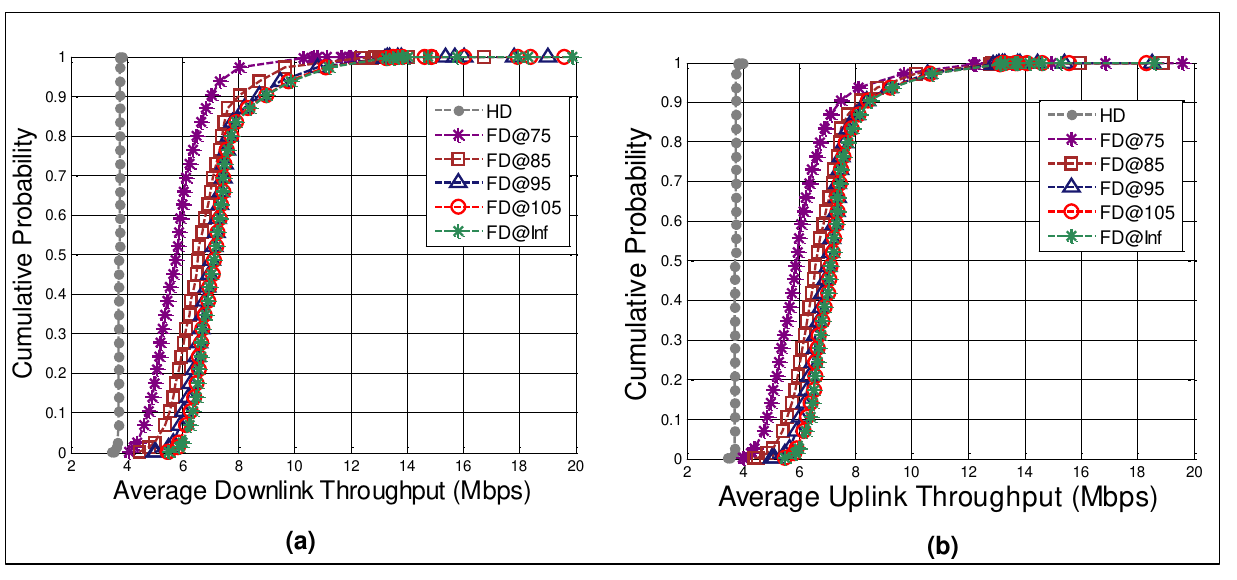}
\caption{Distribution of average data rates for the half-duplex system and full-duplex system with proposed joint UE selection and power allocation in indoor multi-cell scenario.}
\label{fig:fig7}
\vspace{-5mm}
\end{figure}

Next, we generate results with the proposed joint UE selection and power allocation procedure given in Section~\ref{sec3}. Figures~\ref{fig:fig7}(a) and ~\ref{fig:fig7}(b) show the distribution of average downlink and uplink throughputs. Table~\ref{tab2} shows the average throughput gain of the FD system compared to the HD system, and as one would expect, the gain increases as the self-interference cancellation improves. With the higher self-interference cancellation values, the FD system nearly doubles the capacity compared to the HD system. Further, Table~\ref{tab3} shows the average improvement in the 5$\%$ cell edge capacity, which also increases as the self-interference cancellation increases.
\begin {table}
\caption {Average throughput gain of full duplex system over half duplex system in indoor multi-cell scenario.} \label{tab2} 
\vspace{-5mm}
\begin{center}
    \begin{tabular}{| l | l | l | l | l | l |}
    	\hline
		 & \textbf{FD@75} & \textbf{FD@85}  & \textbf{FD@95}  & \textbf{FD@105} & \textbf{FD@Inf} \\ \hline
		\textbf {Downlink} & 56$\%$ & 80$\%$ & 94$\%$ & 97$\%$ & 98$\%$ \\ \hline
		\textbf {Uplink}  & 63$\%$ & 83$\%$ & 93$\%$ & 96$\%$ & 97$\%$\\ \hline		
    \end{tabular}
\end{center}
\end{table}
\begin {table}
\caption {Average improvement in the 5$\%$ cell edge capacity in indoor multi-cell scenario.} \label{tab3} 
\vspace{-5mm}
\begin{center}
    \begin{tabular}{| l | l | l | l | l | l |}
    	\hline
		 & \textbf{FD@75} & \textbf{FD@85}  & \textbf{FD@95}  & \textbf{FD@105} & \textbf{FD@Inf} \\ \hline
		\textbf {Downlink} & 49$\%$ & 74$\%$ & 84$\%$ & 86$\%$ & 87$\%$ \\ \hline
		\textbf {Uplink}  & 55$\%$ & 78$\%$ & 90$\%$ & 93$\%$ & 94$\%$ \\ \hline		
    \end{tabular}
\end{center}
\end{table}

From the simulation one can also observe the dependency between FD/HD operation selection in our scheduler and the self-interference cancellation capability, that is, the lower the self-interference cancellation, the fewer the number of cells in a timeslot that are scheduled in FD mode. This is verified by counting the average number of cells per timeslot which are in FD mode or HD mode or with no transmission as shown in Table~\ref{tab4}. With $75 dB$ self-interference cancellation, on average 84$\%$ of the cells operate in FD mode, while with 105 dB, 98$\%$ of the cells operate in FD mode. In the HD system, in each timeslot, all cells transmit in one direction (either uplink or downlink).
\begin{table}
\caption {Average number of cells per slot in different modes in indoor multi-cell scenario.} \label{tab4} 
\vspace{-5mm}
\begin{center}
\begin{tabular}{|c|c|c|c|c|c|c|}
\hline
\multicolumn{1}{|l|}{} & \begin{tabular}[c]{@{}c@{}}\textbf{HD}\\ (\textbf{Downlink}, \textbf{Uplink})\end{tabular} & \textbf{FD@75} & \textbf{FD@85} & \textbf{FD@95} & \textbf{FD@105} & \textbf{FD@Inf} \\ \hline
\textbf{FD Mode}                & -                                                      & 84\%  & 93\%  & 97\%  & 98\%   & 98\%   \\ \hline
\textbf{HD Mode}                & (100\%, 100\%)                                                  & 16\%  & 7\%   & 3\%   & 2\%    & 2\%    \\ \hline
\textbf{No Transmission}        & (0\%, 0\%)                                                      & 0\%   & 0\%   & 0\%   & 0\%    & 0\%    \\ \hline
\end{tabular}
\end{center}
\end{table}

As energy efficiency becomes a more important performance indicator in future cellular system, we next examine how efficiently the energy is used in both HD and FD operation in terms of \emph{bits/joule}. To calculate this, we keep track of the total throughput and the total transmission power consumed for each UE. The results are shown in Table~\ref{tab5} where we see that there is a penalty in energy efficiency for FD operation that can be quite severe. As the self-interference cancellation improves, the number of UEs transmitting in FD mode increases, resulting in higher inter-node interference, while self-interference reduces. Given this trade-off, the relation between energy efficiency and self-interference cancellation is quite complex. In this scenario, we observe that while the energy efficiency of FD mode can be improved with higher self-interference cancellation, it is still much worse than that of the HD mode. 
\begin {table}
\vspace{-8mm}
\caption {Average energy efficiency in Tbits/joule in indoor multi-cell scenario.} \label{tab5} 
\vspace{-5mm}
\begin{center}
    \begin{tabular}{| l | l | l | l | l | l | l |}
    	\hline
		 & \textbf{HD} & \textbf{FD@75} & \textbf{FD@85}  & \textbf{FD@95}  & \textbf{FD@105} & \textbf{FD@Inf} \\ \hline
		\textbf {Downlink} & 3.74 & 0.045 & 0.097 & 0.227 & 0.326 & 0.434 \\ \hline
		\textbf {Uplink}  & 4.91 & 0.017 & 0.151 & 0.734 & 1.360 & 1.971 \\ \hline		
    \end{tabular}
\end{center}
\end{table}

Since the main reason for the lower energy efficiency of the FD system is the additional power to combat the extra interference, two kinds of solutions can be proposed to alleviate this issue. The first solution is to use techniques to cancel or mitigate the additional interference. The first solution is to cancel or mitigate the additional interference using techniques such as beamforming and sectorization.  In this particular small cell indoor scenario, where most of the inter-cell interference is mitigated by penetration loss between the cells, intra-cell interference plays a dominant role during FD operation. Given that sufficient self-interference cancellation is available for the small cell scenario (e.g., 105 dB), allowing FD operation on the UEs (FD UEs) may remove UE to UE intra-cell interference. In this case, the BS and one UE in each cell will simultaneously transmit in both uplink and downlink directions. Thus, a downlink UE will not experience intra-cell interference from an uplink UE in the same cell. To investigate this observation, we ran our simulation with FD UEs and computed the throughput and energy efficiency. In this case, average throughput gains in the FD system are 44$\%$, 77$\%$, 90$\%$, 99$\%$, and 100$\%$ in the downlink and 43$\%$, 77$\%$, 90$\%$, 99$\%$, and 100$\%$ in the uplink for 75 dB, 85 dB, 95 dB, 105 dB, and perfect self-interference cancellation, respectively.  The average energy efficiency of different systems are shown in Table~\ref{tab5EE1}. For the lower self-interference cancellation case of 75 dB, although the energy efficiency is higher as compared to the previous case of HD UEs, the throughput is lower. As cancellation improves, there is not much difference in the average throughput from the previous case, but energy efficiency improves significantly. In the downlink, 3.04 Tbits/joule is achieved as compared to the 0.326 Tbits/joule and in the uplink, 2.66 Tbits/joule is achieved as compared to the 1.36 Tbits/joule. These results show that in the higher self-interference cancellation scenario, it is beneficial to have FD UEs, especially in a small indoor environment. In this case, energy efficiency does not have monotonic behavior with the self-interference cancellation because of the trade-off mentioned earlier in this section. 
\begin {table}
\caption {Average energy efficiency in Tbits/joule in indoor multi-cell scenario with FD UEs.} \label{tab5EE1} 
\vspace{-5mm}
\begin{center}
    \begin{tabular}{| l | l | l | l | l | l | l |}
    	\hline
		 & \textbf{FD@75} & \textbf{FD@85}  & \textbf{FD@95}  & \textbf{FD@105} & \textbf{FD@Inf} \\ \hline
		\textbf {Downlink} & 0.51 & 2.18 & 1.59 & 3.04 & 3.98 \\ \hline
		\textbf {Uplink}  & 0.31 & 1.60 & 0.86 & 2.66 & 4.08 \\ \hline		
    \end{tabular}
\end{center}
\end{table}

A second solution to improve energy efficiency is to keep using HD UEs but implement a more intelligent scheduling algorithm in which, during the rate/power allocation step, a utility function incorporating the cost of using high power is considered. An example is given in (\ref{20}), 

\begin{equation}\label{20}
\begin{split}
    & \boldsymbol{P}^*(t) = \argmax_{\boldsymbol{P}(t)}  \sum_{b=1}^B [w_{b,\psi_b^d(t)} R^d_{b,\psi_b^d(t)}(t) +  w_{b,\psi_b^u(t)} R^u_{b,\psi_b^u(t)}(t)] - \sum_{b=1}^B  [c_{b,\psi_b^d(t)} f(p^d_{b}(t)) + c_{b,\psi_b^u(t)} f(p^u_{b}(t))]\\
    &\mbox{subject to:} \\[-10pt]
    & \ \ \ \ \ \ \ \ 0 \le p^d_{b}(t) \le P^{d,max},\\[-10pt]
    & \ \ \ \ \ \ \ \ 0 \le p^u_{b}(t) \le P^{u,max}, \  b = \{1,2,...,B\}.
\end{split}
\end{equation} 

The first term is for the capacity maximization same as the given in Section~\ref{sec3b} for the selected UEs. The second term is to take into account power consumption, where $f(p^d_{i}(t))$, and $f(p^u_{i}(t))$ are the functions of power to be allocated to the selected UE in cell $i$ in the downlink and in the uplink direction, respectively. In this term, $c_{i,\psi_i^d(t)}$, and $c_{i,\psi_i^u(t)}$ are the weights to these power terms. In our simulation, $f(.)$ is a logarithmic function of the power. A key parameter in the above formulation is the value of $c_{(.)}$, which impacts the penalty when a UE uses high power. These costs vary for different UEs, for example, UEs further from the cell center should have a lower penalty for high power than UEs nearer to the center.  We use a function of the distance of the UE from its BS, i.e., inversely proportional to the distance of the UE. With such an optimization, average throughput gains in the FD system are 44$\%$, 72$\%$, 91$\%$, 95$\%$, and 96$\%$ in the downlink and 50$\%$, 69$\%$, 85$\%$, 89$\%$, and 91$\%$ in the uplink for 75 dB, 85 dB, 95 dB, 105 dB, and perfect self-interference cancellation, respectively. In this case, we get less throughput gain as compared to the original case, where we did not consider power consumption during the power allocation, but gain a significant improvement in energy efficiency as shown in Table~\ref{tab5EE2}. For example, an energy efficiency of 2.02 Tbits/joule is achieved, compared to the 0.045 Tbits/joule in the downlink with 75 dB SIC. So the scheduler that penalizes high power in the optimization process provides a significant improvement in the energy efficiency for a modest cost in capacity.
\begin {table}
\vspace{-5mm}
\caption {Average energy efficiency in Tbits/joule in indoor multi-cell scenario with power allocation method including penalty to higher power consumption.} \label{tab5EE2} 
\vspace{-5mm}
\begin{center}
    \begin{tabular}{| l | l | l | l | l | l | l |}
    	\hline
		 & \textbf{FD@75} & \textbf{FD@85}  & \textbf{FD@95}  & \textbf{FD@105} & \textbf{FD@Inf} \\ \hline
		\textbf {Downlink} & 2.02 & 1.01 & 0.80 & 0.75 & 0.73 \\ \hline
		\textbf {Uplink}  & 1.46 & 2.47 & 3.23 & 3.58 & 3.61 \\ \hline		
    \end{tabular}
\end{center}
\end{table}

\subsection{Simulation results for sparse outdoor multi-cell environment}\label{sec4b}
The sparse outdoor multi-cell scenario with twelve Pico cells as shown in Figure~\ref{fig:fig5}(b) is investigated in this section. The simulation parameters are based on 3GPP simulation recommendations for outdoor Pico cells~\cite{3GPP:1}, and are described in Table~\ref{tab6}. The probability of LOS for BS-to-BS and BS-to-UE path loss is (R is in kilometers),
\begin{equation}\label{21}
P_{LOS} = 0.5 - min(0.5, 5exp(-0.156/R)) + min(0.5, 5exp(-R/0.03)).
\end{equation}

\begin {table}
\caption {Simulation parameters for outdoor multi-cell scenario} \label{tab6} 
\vspace{-5mm}
\begin{center}
    \begin{tabular}{| p{2 in} | p{4 in} |}
    	\hline
		\textbf {Parameter} & \textbf{Value} \\ \hline
		Maximum BS Power & 24 dBm \\ \hline
		Maximum UE Power & 23 dBm \\ \hline
		Minimum distance between Pico BSs & 40 m \\ \hline
		Radius of a Pico cell & 40 m \\ \hline
		Thermal Noise Density & -174 dBm/Hz \\ \hline
		Noise Figure & BS: 13 dB, UE: 9 dB \\ \hline
		Shadowing standard deviation between BS and UE &  LOS:~ $3~dB$ NLOS: $4~dB$ \\  \hline
		Shadowing standard deviation between Pico cells &  $6~dB$ \\  \hline
		BS-to-BS pathloss (R in kilometers) & LOS: if $R < 2/3 km, PL(R) = 98.4 + 20~log_{10}(R)$, else $PL(R) = 101.9 + 40~log_{10}(R)$, NLOS: $PL(R) = 169.36 + 40 log_{10}(R)$.  \\ \hline
		BS-to-UE pathloss (R in kilometers) & LOS: $PL(R) = 103.8 + 20.9~log_{10}(R),$ NLOS: $PL(R) = 145.4 + 37.5~log_{10}(R)$.  \\ \hline
		UE-to-UE pathloss (R in kilometers) & If $R \leq 50 m, PL(R) = 98.45 + 20~log_{10}(R),$ else,  $PL(R) = 175.78 + 40~log_{10}(R)$.  \\ \hline
    \end{tabular}
\end{center}
\end{table}

Ten randomly distributed UEs are deployed in each cell. With these settings, we run our simulation for several random drops of twelve Pico cells in the given area of a hexagonal cell with height of 500 meters. We generate the results with the proposed joint UE selection and power allocation method given in Section~\ref{sec3}. 
\begin{figure}
\centering
\includegraphics[width = 5in] {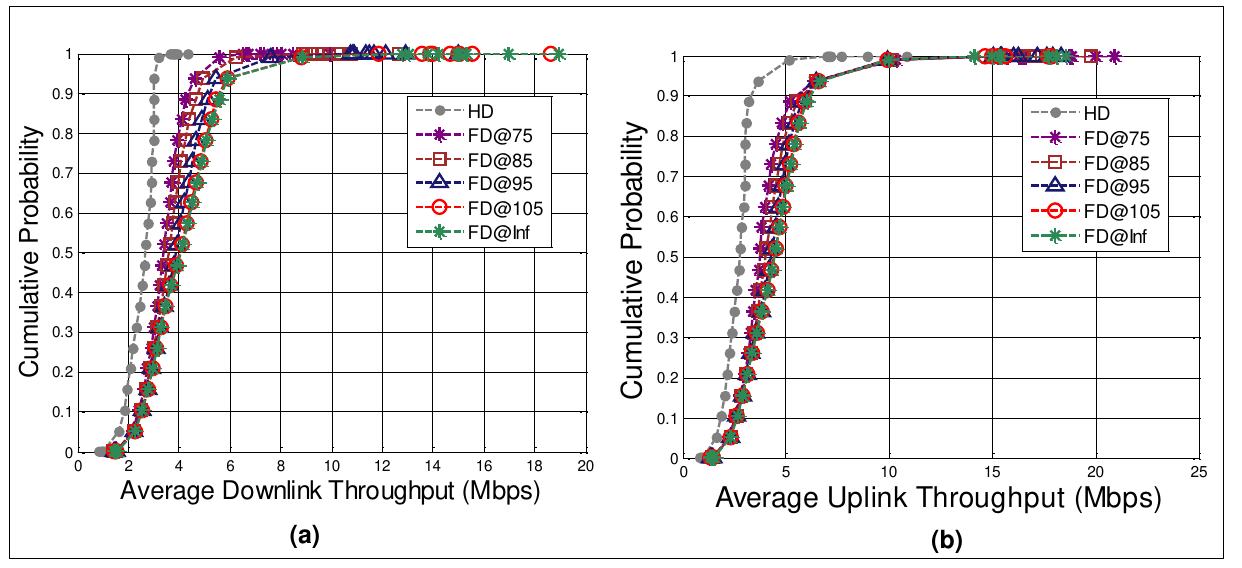}
\caption{Distribution of average data rates for the half-duplex system and full-duplex system with proposed joint UE selection and power allocation in outdoor multi-cell scenario.}
\label{fig:fig8}
\end{figure}

\begin {table}
\caption {Average throughput gain of full duplex system over half duplex system in outdoor multi-cell scenario.} \label{tab7} 
\vspace{-5mm}
\begin{center}
    \begin{tabular}{| l | l | l | l | l | l |}
    	\hline
		 & \textbf{FD@75} & \textbf{FD@85}  & \textbf{FD@95}  & \textbf{FD@105} & \textbf{FD@Inf} \\ \hline
		\textbf {Downlink} & 34$\%$ & 42$\%$ & 53$\%$ & 60$\%$ & 62$\%$ \\ \hline
		\textbf {Uplink}  & 47$\%$ & 54$\%$ & 60$\%$ & 63$\%$ & 64$\%$\\ \hline		
    \end{tabular}
\end{center}
\end{table}

Figures~\ref{8}(a) and~\ref{8}(b) show the distribution of average downlink and uplink throughputs, and Table~\ref{tab7} shows the average throughput gain of the FD system compared to the HD system. Similar to the previous scenario, FD increases the capacity of the system significantly over the HD case, where the increase is proportional to the amount of self-interference cancellation. In this outdoor scenario, the average throughput of a UE is lower compared to the indoor case but it is distributed over a wider range. Moreover, the throughput increase due to FD operation is less than what it was in the indoor case. The reason behind this is that the inter-cell interference between a BS and UEs in neighboring cells is much stronger that in the indoor scenario.

Table~\ref{tab8} shows the average number of cells per slot which are in FD mode, HD mode or with no transmission. First of all, in the HD system, contrary to the indoor scenario, we can see that some cells are not transmitting. This is due to the higher inter-cell interference between the BS and UEs in neighboring cells; the system throughput is higher when certain cells are not scheduled for transmission, resulting in reduced inter-cell interference. Further, for the same reason, the average number of cells operating in FD mode is smaller than the indoor scenario. In this case, the number of cells in FD mode also increases with self-interference cancellation.
\begin{table}
\caption {Average number of cells per slot in different modes in outdoor multi-cell scenario.} \label{tab8}
\vspace{-5mm}
\begin{center}
\begin{tabular}{|c|c|c|c|c|c|c|}
\hline
\multicolumn{1}{|l|}{} & \begin{tabular}[c]{@{}c@{}}\textbf{HD}\\ (\textbf{Downlink}, \textbf{Uplink})\end{tabular} & \textbf{FD@75} & \textbf{FD@85} & \textbf{FD@95} & \textbf{FD@105} & \textbf{FD@Inf} \\ \hline
\textbf{FD Mode}                & -                                                    & 36\%  & 50\%  & 56\%  & 57\%   & 57\%   \\ \hline
\textbf{HD Mode}                & (91\%, 98\%)                                                  & 62\%  & 48\%   & 42\%   & 41\%    & 41\%    \\ \hline
\textbf{No Transmission}        & (9\%, 2\%)                                                      & 2\%   & 2\%   & 2\%   & 2\%    & 2\%    \\ \hline
\end{tabular}
\end{center}
\end{table}

Table~\ref{tab9} shows the average energy efficiency results for both HD and FD operation in terms of bits/joule. Note that in this outdoor scenario, for most of the cases (except FD@75 downlink), energy efficiency is lower than the previous indoor case. This is again due to the high inter-cell interference between a BS and UEs in neighboring cells. For the FD@75 downlink case, energy efficiency is even higher than the HD case. This is because in an FD system, a downlink UE suffers interference from uplink UEs of neighboring cells and/or BSs of neighboring cells. It is observed in our simulations that in general, UE to UE interference is lower than the BS to UE interference. In case of FD@75, for most of the cells ($\sim$62$\%$) there is only one transmission, with 23$\%$ of cells in downlink and 39$\%$ in uplink. Thus, since UE to UE inter-cell interference is lower, we get higher energy efficiency in downlink of FD@75 compared to the downlink in HD where a downlink UE gets BS to UE interference from all of its active neighboring cells. Further, as the self-interference cancellation increases, energy efficiency is decreased due to a higher number of simultaneous transmissions. Also, as described in the Section IV.A, an increase in self-interference cancellation may not always guarantee a reduction in interference for a UE in the FD system. Due to this trade-off, uplink energy efficiency first decreases, then further increases with self-interference cancellation.

\begin {table}
\caption {Average energy efficiency in Tbits/joule in outdoor multi-cell scenario.} \label{tab9} 
\vspace{-5mm}
\begin{center}
    \begin{tabular}{| l | l | l | l | l | l | l |}
    	\hline
		 & \textbf{HD} & \textbf{FD@75} & \textbf{FD@85}  & \textbf{FD@95}  & \textbf{FD@105} & \textbf{FD@Inf} \\ \hline
		\textbf {Downlink} & 0.07 & 0.15 & 0.046 & 0.026 & 0.023 & 0.023 \\ \hline
		\textbf {Uplink}  & 0.017 & 0.007 & 0.003 & 0.005 & 0.012 & 0.016 \\ \hline		
    \end{tabular}
\end{center}
\end{table}

In this paper, symmetric traffic demands in uplink and downlink are considered. We acknowledge that asymmetric traffic demands will reduce the need for simultaneous uplink and downlink transmission. This will decrease the potential capacity gain, which can be achieved by FD operation. However, recent trend in online storage services like Dropbox, Google Drive, iCloud, etc., and increasing popularity in uploading of videos and photos to social networking sites will continue to increase the uplink traffic volume significantly, which will make uplink and downlink traffic less asymmetric. As mentioned in Section~\ref{sec3}, we considered a centralized scheduler in this paper. We acknowledge that cooperation efforts among base stations are complex, and the complexity increases with the number of cells and number of UEs in each cell. Performance could also degrade with latency in the feedback of CSI. However, given the recent developments in coordinated multi-point transmission (CoMP) and cloud RAN (C-RAN) technologies, such cooperation could become more practical in the near future. We are currently working on distributed and semi-distributed scheduling algorithms for the multi-cell FD system.

\section{Conclusion}\label{sec5}
We investigated the application of common carrier FD radios to resource managed small-cell systems in a multi-cell deployment. Assuming FD capable base stations with legacy user equipment, a joint scheduling and power allocation method was proposed, which can apply to both HD and FD systems. In the FD system, it operates in FD mode when conditions are favorable, and otherwise defaults to HD mode. We evaluate the performance of our scheduler in both indoor and outdoor multi-cell environments. Our simulation results show that an FD system using a practical design parameter of 95 dB self-interference cancellation at each base station can improve the capacity by 94$\%$ in the downlink and 93$\%$ in the uplink in an indoor multi-cell hot zone scenario and 53$\%$ in the downlink and 60$\%$ in the uplink in an outdoor multi Pico cell scenario. From these results we conclude that in both indoor small-cell and sparse outdoor environment, FD base stations with an intelligent scheduling algorithm are able to improve capacity significantly. We observed a penalty in energy efficiency during FD operation. Further, we discussed the ways to increase the energy efficiency of FD system by enabling FD UEs, specially in a small indoor environment, or using a modified scheduling algorithm that penalizes using high power during the FD operation. We continue to investigate FD resource management algorithms with manageable complexity and information exchange requirements, that incorporate energy efficiency as a performance metric, and that provide performance improvement consistent with the promising results achieved so far.

\bibliographystyle{IEEEtran}
\bibliography{FD_references}

@MISC{FCC,
        AUTHOR =        {{Federal Communications Commission}},
        TITLE =         {{Mobile Broadband: The Benefits of Additional Spectrum}},
        Month = {October},
        Year =          {2010},
        Note =          {},
        url = {www.fcc.gov},
        summary =       {}
        }
        
@MISC{Cisco,
        AUTHOR =        {Cisco},
        TITLE =         {{Cisco visual network index}: Forecast and methodology 2012-2017 },
        Howpublished =  {Cisco white paper},
        Date = {29},
        Month = {May},
        Year =          {2013},
        Note =          {},
        url = {www.cisco.com},
        summary =       {}
        }
        
@MISC{Ericsson,
        AUTHOR =        {Ericsson},
        TITLE =         {Ericsson mobility report},
         Month = {June},
        Year =          {2013},
        Note =          {},
        url = {www.ericsson.com},
        summary =       {}
        }

@inproceedings{difazio2011bandwidth,
  title={The bandwidth crunch: Can wireless technology meet the skyrocketing demand for mobile data?},
  author={DiFazio, Robert A and Pietraski, Philip J},
  booktitle={Proc. Long Island Systems, Applications and Technology Conference (LISAT)},
  year={2011},
  organization={IEEE}
}   

@MISC{Horizon,
        AUTHOR =        {},
        TITLE =         {Creating a smart network that is flexible, robust and cost effective},
        Howpublished =  {Horizon 2020 Advanced 5G Network Infrastructure
for Future Internet PPP, Industry Proposal (Draft Version 2.1)},
         Year =          {2013},
        Note =          {},
        url = {http://www.networks-etp.eu/},
        summary =       {}
        }  
        
 @MISC{EB,
        AUTHOR =        {{Electrical Business}},
        TITLE =         {Searching for 1000 times the capacity of 4G wireless},
        
        Month = {July},
         Year =          {2013},
        Note =          {},
        url = {www.ebmag.com/Industry-News/searching-for-1000-times-the-capacity-of-4g-wireless.html},
        summary =       {}
        }  
        
  @MISC{Qualcomm,
        AUTHOR =        {Qualcomm},
        TITLE =         {The 1000x mobile data challenge},
         Year =          {2012},
        Note =          {},
        url = {www.qualcomm.com/1000x},
        summary =       {}
        }  
        
@misc{Khandani10,
  title={Methods for spatial multiplexing of wireless two-way channels},
  author={Khandani, Amir Keyvan},
  year={2010},
  month=oct # "~19",
  publisher={Google Patents},
  note={{US} Patent 7,817,641}
}        

@INPROCEEDINGS{Katti10,
        AUTHOR =     "J. Choi and M. Jain and K. Srinivasan and P. Levis and S. Katti",
        TITLE =      "Achieving single channel, full duplex wireless communication",
        BOOKTITLE    = "Proc. Sixteenth Annual Intl Conf. on Mobile Computing and Networking ({MobiCom})",
        YEAR    =   {2010}
}        

@INPROCEEDINGS{Knox12,
        AUTHOR =     "M. Knox",
        TITLE =      "Single antenna full duplex communications using a common carrier",
        BOOKTITLE    = "Proc. 13th IEEE Annual Wireless and Microwave Technology Conference (WAMICON)",
        YEAR    =   {2012}
}

@INPROCEEDINGS{Katti13,
        AUTHOR =     "D. Bharadia and E. McMilin and S. Katti",
        TITLE =      "Full duplex radios",
        BOOKTITLE    = "Proc. ACM SIGCOMM 2013",
        YEAR    =   {2013}
}

@article{ Duarte13,
   author = "M. Duarte and A. Sabharwal and V. Aggarwal and  R. Jana and K. Ramakrishnan and C. Rice and N. Shankaranarayanan",
   title = {Design and Characterization of a Full-duplex Multi-antenna System for {WiFi} networks},
   journal = "IEEE Trans. Vehicular Technologies",
   year =     {2013}
   }    
  
@article{Sahai11,
  author    = {Achaleshwar Sahai and
               Gaurav Patel and
               Ashutosh Sabharwal},
  title     = {Pushing the limits of Full-duplex: Design and Real-time Implementation},
  journal   = {CoRR},
  volume    = {abs/1107.0607},
  year      = {2011},
  url       = {http://arxiv.org/abs/1107.0607},
  timestamp = {Mon, 05 Dec 2011 18:05:27 +0100},
  biburl    = {http://dblp.uni-trier.de/rec/bib/journals/corr/abs-1107-0607},
  bibsource = {dblp computer science bibliography, http://dblp.org}
}

@INPROCEEDINGS{SanjayAsilomar13, 
author={Sanjay Goyal and Pei Liu and Ozgur Gurbuz and  Elza Erkip and Shivendra Panwar}, 
booktitle={Forty Seventh Asilomar Conference on Signals, Systems and Computers (ASILOMAR)}, 
title={A Distributed {MAC} Protocol for Full Duplex Radio}, 
year={2013}, 
}

@INPROCEEDINGS{Singh11, 
author={Singh, N. and Gunawardena, D. and Proutiere, A. and Radunovic, B. and Balan, H.V. and Key, P.}, 
booktitle={International Symposium on Modeling and Optimization in Mobile, Ad Hoc and Wireless Networks (WiOpt)}, 
title={Efficient and fair {MAC} for wireless networks with self-interference cancellation}, 
year={2011}, 
pages={94-101}, 
}
@INPROCEEDINGS{Barghi12,
        AUTHOR =     "S. Barghi and A. Khojastepour and K. Sundaresan and S. Rangarajan",
        TITLE =      "Characterizing the throughput gain of single cell {MIMO} wireless systems with full duplex radios",
        BOOKTITLE    = "Proc. Intl. Symposium on Modeling and Optimization in Mobile, Ad Hoc, and Wireless Networks(WiOpt)",
        YEAR    =   {2012}
}

@INPROCEEDINGS{XShen13, 
author={Xia Shen and Xiang Cheng and Liuqing Yang and Meng Ma and Bingli Jiao}, 
booktitle={IEEE Global Communications Conference (GLOBECOM)}, 
title={On the Design of the Scheduling Algorithm for the Full Duplexing Wireless Cellular Network}, 
year={2013}, 
}

@article{Simeone2014full,
  title={Full-Duplex Cloud Radio Access Networks: An Information-Theoretic Viewpoint},
  author={Simeone, Osvaldo and Erkip, Elza and Shamai, Shlomo},
  journal={arXiv preprint arXiv:1405.2092},
  year={2014}
}

@ARTICLE{ChoiSTR12, 
author={Choi, Yang-Seok and Shirani-Mehr, Hooman}, 
journal={IEEE Transactions on Wireless Communications }, 
title={Simultaneous Transmission and Reception: Algorithm, Design and System Level Performance}, 
year={2013}, 
volume={12}, 
number={12}, 
pages={5992-6010}, 
}

@INPROCEEDINGS{SanjayCISS13,
        AUTHOR =     "S. Goyal and P. Liu and S. Hua and S. Panwar",
        TITLE =      "Analyzing a full-duplex cellular system",
        booktitle={Proc. 44th Annual Conference on Information Sciences and Systems (CISS)},
        Month = {March},
        YEAR    =   {2013}
}

@MISC{3GPP:1,
        AUTHOR =        {{3GPP}},
        TITLE =         {Further enhancements to LTE Time Division Duplex ({TDD}) for Downlink-Uplink ({DL-UL})
interference management and traffic adaptation },
        Howpublished =  {TR 36.828, v.11.0.0},
        Month = {June},
        Year =          {2012},
        Note =          {},
        url = {www.3gpp.org},
        summary =       {}
        }
        
 @INPROCEEDINGS{SanjayICC14,
        AUTHOR =     "S. Goyal and P. Liu and S. Panwar and R. DiFazio and R. Yang and J. Li and E. Bala",
        TITLE =      "Improving small cell capacity with common-carrier full duplex radios",
        BOOKTITLE    = "IEEE International Conference on Communications (ICC)",
        Month = {June},
        YEAR    =   {2014}
}

@BOOK{DahlmanLTE,
        AUTHOR =        {E. Dahlman and S. Parkvall and J. Skold},
        editor =        {},
        TITLE =         {{4G LTE / LTE-Advanced for Mobile Broadband}},
        Volume =        {},
        Edition =       {},
        Series =        {},
        PUBLISHER =     {Oxford:Elsevier},
        Address =       {},
        Month =         {},
        YEAR =          2011,
        Note =          {},
        summary =       {}
        }  

@MISC{3GPP:2,
        AUTHOR =        {{3GPP}},
        TITLE =         {Physical Channels and Modulation ({Release 10})},
        Howpublished =  {TS 36.211, v.10.5.0 },
        Month = {June},
        Year =          {2012},
        Note =          {},
        url = {www.3gpp.org},
        summary =       {}
        }
@MISC{3GPP:3,
        AUTHOR =        {{3GPP}},
        TITLE =         {Further advancements for E-UTRA physical layer aspects ({Release 9})},
        Howpublished =  {TR 36.814, v.9.0.0},
        Month = {March},
        Year =          {2010},
        Note =          {},
        url = {www.3gpp.org},
        summary =       {}
        }
@INPROCEEDINGS{Stoylar05,
        AUTHOR =     "A. L. Stoylar",
        TITLE =      "On the Asymptotic Optimality of the Gradient Scheduling Algorithm for Multi-User Throughput
Allocation",
        BOOKTITLE    = "Operation Research",
        YEAR    =   {2005}
}

@inproceedings{jalali2000data,
  title={Data throughput of CDMA-HDR a high efficiency-high data rate personal communication wireless system},
  author={Jalali, A and Padovani, R and Pankaj, R},
  booktitle={Proc. IEEE 51st Vehicular Technology Conference (VTC), 2000},
  year={2000},
  organization={IEEE}
}
@article{girici2010proportional,
  title={Proportional fair scheduling algorithm in OFDMA-based wireless systems with {QoS} constraints},
  author={Girici, Tolga and Zhu, Chenxi and Agre, Jonathan R and Ephremides, Anthony},
  journal={Journal of Communications and Networks},
  volume={12},
  number={1},
  pages={30--42},
  year={2010},
  publisher={IEEE}
}

@article{venturino2009coordinated,
  title={Coordinated scheduling and power allocation in downlink multicell {OFDMA} networks},
  author={Venturino, Luca and Prasad, Narayan and Wang, Xiaodong},
  journal={IEEE Transactions on Vehicular Technology},
  volume={58},
  number={6},
  pages={2835--2848},
  year={2009},
  publisher={IEEE}
}

@inproceedings{yu2010joint,
  title={Joint scheduling and dynamic power spectrum optimization for wireless multicell networks},
  author={Yu, Wei and Kwon, Taesoo and Shin, Changyong},
  booktitle={Proc. 44th Annual Conference onInformation Sciences and Systems (CISS)},
  pages={1--6},
  year={2010},
  organization={IEEE}
}


@inproceedings{kiani2007maximizing,
  title={Maximizing multicell capacity using distributed power allocation and scheduling},
  author={Kiani, Saad G and Oien, Geir E and Gesbert, David},
  booktitle={IEEE Wireless Communications and Networking Conference (WCNC)},
  pages={1690--1694},
  year={2007},
  organization={IEEE}
}

@article{koutsopoulos2006cross,
  title={Cross-layer adaptive techniques for throughput enhancement in wireless {OFDM}-based networks},
  author={Koutsopoulos, Iordanis and Tassiulas, Leandros},
  journal={IEEE/ACM Transactions on Networking (TON)},
  volume={14},
  number={5},
  pages={1056--1066},
  year={2006},
  publisher={IEEE Press}
}

@article{boyd2007tutorial,
  title={A tutorial on geometric programming},
  author={Boyd, Stephen and Kim, Seung Jean and Vandenberghe, Lieven and Hassibi, Arash},
  journal={Optimization and engineering},
  volume={8},
  number={1},
  pages={67--127},
  year={2007},
  publisher={Springer}
}

@article{chiang2007power,
  title={Power control by geometric programming},
  author={Chiang, Mung and Tan, Chee Wei and Palomar, Daniel P and O'Neill, Daniel and Julian, David},
  journal={Wireless Communications, IEEE Transactions on},
  volume={6},
  number={7},
  pages={2640--2651},
  year={2007},
  publisher={IEEE}
}


\begin{thebibliography}{10}
\providecommand{\url}[1]{#1}
\csname url@samestyle\endcsname
\providecommand{\newblock}{\relax}
\providecommand{\bibinfo}[2]{#2}
\providecommand{\BIBentrySTDinterwordspacing}{\spaceskip=0pt\relax}
\providecommand{\BIBentryALTinterwordstretchfactor}{4}
\providecommand{\BIBentryALTinterwordspacing}{\spaceskip=\fontdimen2\font plus
\BIBentryALTinterwordstretchfactor\fontdimen3\font minus
  \fontdimen4\font\relax}
\providecommand{\BIBforeignlanguage}[2]{{%
\expandafter\ifx\csname l@#1\endcsname\relax
\typeout{** WARNING: IEEEtran.bst: No hyphenation pattern has been}%
\typeout{** loaded for the language `#1'. Using the pattern for}%
\typeout{** the default language instead.}%
\else
\language=\csname l@#1\endcsname
\fi
#2}}
\providecommand{\BIBdecl}{\relax}
\BIBdecl

\bibitem{FCC}
\BIBentryALTinterwordspacing
{Federal Communications Commission}, ``{Mobile Broadband: The Benefits of
  Additional Spectrum},'' October 2010. [Online]. Available: \url{www.fcc.gov}
\BIBentrySTDinterwordspacing

\bibitem{Cisco}
\BIBentryALTinterwordspacing
Cisco, ``{Cisco visual network index}: Forecast and methodology 2013-2018,''
  Cisco white paper, June 2014. [Online]. Available: \url{www.cisco.com}
\BIBentrySTDinterwordspacing

\bibitem{Ericsson}
\BIBentryALTinterwordspacing
Ericsson, ``Ericsson mobility report,'' June 2013. [Online]. Available:
  \url{www.ericsson.com}
\BIBentrySTDinterwordspacing

\bibitem{difazio2011bandwidth}
R.~A. DiFazio and P.~J. Pietraski, ``The bandwidth crunch: Can wireless
  technology meet the skyrocketing demand for mobile data?'' in \emph{Proc.
  Long Island Systems, Applications and Technology Conference (LISAT)}.\hskip
  1em plus 0.5em minus 0.4em\relax IEEE, May 2011.

\bibitem{Horizon}
\BIBentryALTinterwordspacing
``Creating a smart network that is flexible, robust and cost effective,''
  Horizon 2020 Advanced 5G Network Infrastructure for Future Internet PPP,
  Industry Proposal (Draft Version 2.1), 2013. [Online]. Available:
  \url{http://www.networks-etp.eu/}
\BIBentrySTDinterwordspacing

\bibitem{EB}
\BIBentryALTinterwordspacing
{Electrical Business}, ``Searching for 1000 times the capacity of 4{G}
  wireless,'' July 2013. [Online]. Available:
  \url{www.ebmag.com/Industry-News/searching-for-1000-times-the-capacity-of-4G-wireless.html}
\BIBentrySTDinterwordspacing

\bibitem{Qualcomm}
\BIBentryALTinterwordspacing
Qualcomm, ``The 1000x mobile data challenge,'' 2012. [Online]. Available:
  \url{www.qualcomm.com/1000x}
\BIBentrySTDinterwordspacing

\bibitem{Khandani10}
A.~K. Khandani, ``Methods for spatial multiplexing of wireless two-way
  channels,'' October 2010, {US} Patent 7,817,641.

\bibitem{Katti10}
J.~Choi, M.~Jain, K.~Srinivasan, P.~Levis, and S.~Katti, ``Achieving single
  channel, full duplex wireless communication,'' in \emph{Proc. Sixteenth
  Annual Intl Conf. on Mobile Computing and Networking ({MobiCom})}, 2010.

\bibitem{Knox12}
M.~Knox, ``Single antenna full duplex communications using a common carrier,''
  in \emph{Proc. 13th IEEE Annual Wireless and Microwave Technology Conference
  (WAMICON)}, 2012.

\bibitem{Katti13}
D.~Bharadia, E.~McMilin, and S.~Katti, ``Full duplex radios,'' in \emph{Proc.
  ACM SIGCOMM 2013}, 2013.

\bibitem{Duarte13}
M.~Duarte, A.~Sabharwal, V.~Aggarwal, R.~Jana, K.~Ramakrishnan, C.~Rice, and
  N.~Shankaranarayanan, ``Design and characterization of a full-duplex
  multi-antenna system for {WiFi} networks,'' \emph{IEEE Trans. Vehicular
  Technologies}, 2013.

\bibitem{Sahai11}
\BIBentryALTinterwordspacing
A.~Sahai, G.~Patel, and A.~Sabharwal, ``Pushing the limits of full-duplex:
  Design and real-time implementation,'' \emph{CoRR}, vol. abs/1107.0607, 2011.
  [Online]. Available: \url{http://arxiv.org/abs/1107.0607}
\BIBentrySTDinterwordspacing

\bibitem{SanjayAsilomar13}
S.~Goyal, P.~Liu, O.~Gurbuz, E.~Erkip, and S.~Panwar, ``A distributed {MAC}
  protocol for full duplex radio,'' in \emph{Forty Seventh Asilomar Conference
  on Signals, Systems and Computers (ASILOMAR)}, 2013.

\bibitem{Singh11}
N.~Singh, D.~Gunawardena, A.~Proutiere, B.~Radunovic, H.~Balan, and P.~Key,
  ``Efficient and fair {MAC} for wireless networks with self-interference
  cancellation,'' in \emph{International Symposium on Modeling and Optimization
  in Mobile, Ad Hoc and Wireless Networks (WiOpt)}, 2011, pp. 94--101.

\bibitem{Barghi12}
S.~Barghi, A.~Khojastepour, K.~Sundaresan, and S.~Rangarajan, ``Characterizing
  the throughput gain of single cell {MIMO} wireless systems with full duplex
  radios,'' in \emph{Proc. Intl. Symposium on Modeling and Optimization in
  Mobile, Ad Hoc, and Wireless Networks (WiOpt)}, 2012.

\bibitem{DiINFOCOM}
B.~Di, S.~Bayat, L.~Song, and Y.~Li, ``Radio resource allocation for
  full-duplex ofdma networks using matching theory,'' in \emph{IEEE Conference
  on Computer Communications Workshops (INFOCOM WKSHPS)}.\hskip 1em plus 0.5em
  minus 0.4em\relax IEEE, 2014, pp. 197--198.

\bibitem{Shaocommletter}
S.~Shao, D.~Liu, K.~Deng, Z.~Pan, and Y.~Tang, ``Analysis of carrier
  utilization in full-duplex cellular networks by dividing the co-channel
  interference region,'' \emph{IEEE Communications Letters}, vol.~18, no.~6,
  pp. 1043--1046, June 2014.

\bibitem{XShen13}
X.~Shen, X.~Cheng, L.~Yang, M.~Ma, and B.~Jiao, ``On the design of the
  scheduling algorithm for the full duplexing wireless cellular network,'' in
  \emph{IEEE Global Communications Conference (GLOBECOM)}, 2013.

\bibitem{Simeone2014full}
O.~Simeone, E.~Erkip, and S.~Shamai, ``Full-duplex cloud radio access networks:
  An information-theoretic viewpoint,'' \emph{arXiv preprint arXiv:1405.2092},
  2014.

\bibitem{HyunICTC}
H.-H. Choi, ``On the design of user pairing algorithms in full duplexing
  wireless cellular networks,'' in \emph{International Conference on
  Information and Communication Technology Convergence (ICTC)}, Oct 2014, pp.
  490--495.

\bibitem{ChoiSTR12}
Y.-S. Choi and H.~Shirani-Mehr, ``Simultaneous transmission and reception:
  Algorithm, design and system level performance,'' \emph{IEEE Transactions on
  Wireless Communications}, vol.~12, no.~12, pp. 5992--6010, 2013.

\bibitem{duplo_site}
\BIBentryALTinterwordspacing
{The Duplo Website}. [Online]. Available: \url{http://www.fp7-duplo.eu/}
\BIBentrySTDinterwordspacing

\bibitem{Nguyenduplo}
D.~Nguyen, L.~Tran, P.~Pirinen, and M.~Latva{-}aho, ``On the spectral
  efficiency of full-duplex small cell wireless systems,'' \emph{CoRR}, vol.
  abs/1407.2628, 2014.

\bibitem{SanjayCISS13}
S.~Goyal, P.~Liu, S.~Hua, and S.~Panwar, ``Analyzing a full-duplex cellular
  system,'' in \emph{Proc. 44th Annual Conference on Information Sciences and
  Systems (CISS)}, March 2013.

\bibitem{alves2014average}
H.~Alves, C.~H. de~Lima, P.~H. Nardelli, R.~Demo~Souza, and M.~Latva-aho, ``On
  the average spectral efficiency of interference-limited full-duplex
  networks,'' in \emph{9th International Conference on Cognitive Radio Oriented
  Wireless Networks and Communications (CROWNCOM)}.\hskip 1em plus 0.5em minus
  0.4em\relax IEEE, 2014, pp. 550--554.

\bibitem{Quekhybrid}
J.~Lee and T.~Q. Quek, ``Hybrid full-/half-duplex system analysis in
  heterogeneous wireless networks,'' \emph{arXiv preprint arXiv:1411.4848},
  2014.

\bibitem{3GPP:1}
\BIBentryALTinterwordspacing
{3GPP}, ``Further enhancements to lte time division duplex ({TDD}) for
  downlink-uplink ({DL-UL}) interference management and traffic adaptation,''
  TR 36.828, v.11.0.0, June 2012. [Online]. Available: \url{www.3gpp.org}
\BIBentrySTDinterwordspacing

\bibitem{SanjayICC14}
S.~Goyal, P.~Liu, S.~Panwar, R.~DiFazio, R.~Yang, J.~Li, and E.~Bala,
  ``Improving small cell capacity with common-carrier full duplex radios,'' in
  \emph{IEEE International Conference on Communications (ICC)}, June 2014.

\bibitem{DahlmanLTE}
E.~Dahlman, S.~Parkvall, and J.~Skold, \emph{{4G LTE / LTE-Advanced for Mobile
  Broadband}}.\hskip 1em plus 0.5em minus 0.4em\relax Oxford:Elsevier, 2011.

\bibitem{3GPP:2}
\BIBentryALTinterwordspacing
{3GPP}, ``Physical channels and modulation ({Release 10}),'' TS 36.211,
  v.10.5.0, June 2012. [Online]. Available: \url{www.3gpp.org}
\BIBentrySTDinterwordspacing

\bibitem{3GPP:3}
\BIBentryALTinterwordspacing
------, ``Further advancements for {E-UTRA} physical layer aspects ({Release
  9}),'' TR 36.814, v.9.0.0, March 2010. [Online]. Available:
  \url{www.3gpp.org}
\BIBentrySTDinterwordspacing

\bibitem{Stoylar05}
A.~L. Stoylar, ``On the asymptotic optimality of the gradient scheduling
  algorithm for multi-user throughput allocation,'' in \emph{Operation
  Research}, 2005.

\bibitem{jalali2000data}
A.~Jalali, R.~Padovani, and R.~Pankaj, ``Data throughput of {CDMA-HDR} a high
  efficiency-high data rate personal communication wireless system,'' in
  \emph{Proc. IEEE 51st Vehicular Technology Conference (VTC), 2000}.\hskip 1em
  plus 0.5em minus 0.4em\relax IEEE, 2000.

\bibitem{girici2010proportional}
T.~Girici, C.~Zhu, J.~R. Agre, and A.~Ephremides, ``Proportional fair
  scheduling algorithm in {OFDMA}-based wireless systems with {QoS}
  constraints,'' \emph{Journal of Communications and Networks}, vol.~12, no.~1,
  pp. 30--42, 2010.

\bibitem{venturino2009coordinated}
L.~Venturino, N.~Prasad, and X.~Wang, ``Coordinated scheduling and power
  allocation in downlink multicell {OFDMA} networks,'' \emph{IEEE Transactions
  on Vehicular Technology}, vol.~58, no.~6, pp. 2835--2848, 2009.

\bibitem{yu2010joint}
W.~Yu, T.~Kwon, and C.~Shin, ``Joint scheduling and dynamic power spectrum
  optimization for wireless multicell networks,'' in \emph{Proc. 44th Annual
  Conference onInformation Sciences and Systems (CISS)}.\hskip 1em plus 0.5em
  minus 0.4em\relax IEEE, 2010, pp. 1--6.

\bibitem{kiani2007maximizing}
S.~G. Kiani, G.~E. Oien, and D.~Gesbert, ``Maximizing multicell capacity using
  distributed power allocation and scheduling,'' in \emph{IEEE Wireless
  Communications and Networking Conference (WCNC)}.\hskip 1em plus 0.5em minus
  0.4em\relax IEEE, 2007, pp. 1690--1694.

\bibitem{koutsopoulos2006cross}
I.~Koutsopoulos and L.~Tassiulas, ``Cross-layer adaptive techniques for
  throughput enhancement in wireless {OFDM}-based networks,'' \emph{IEEE/ACM
  Transactions on Networking (TON)}, vol.~14, no.~5, pp. 1056--1066, 2006.

\bibitem{boyd2007tutorial}
S.~Boyd, S.~J. Kim, L.~Vandenberghe, and A.~Hassibi, ``A tutorial on geometric
  programming,'' \emph{Optimization and engineering}, vol.~8, no.~1, pp.
  67--127, 2007.

\bibitem{chiang2007power}
M.~Chiang, C.~W. Tan, D.~P. Palomar, D.~O'Neill, and D.~Julian, ``Power control
  by geometric programming,'' \emph{IEEE Transactions on Wireless
  Communications}, vol.~6, no.~7, pp. 2640--2651, 2007.

\end{thebibliography}

\end{document}